\documentclass[12pt,letterpaper]{article}
\usepackage[utf8]{inputenc}
\DeclareUnicodeCharacter{03B8}{\circ}
\usepackage{graphicx}
\usepackage{scalerel}
\usepackage{amsmath}
\usepackage{booktabs}
\usepackage{tabu}
\usepackage{graphicx}
\usepackage{caption}
\usepackage{subcaption}
\usepackage{cite}
\usepackage{secdot}
\usepackage[margin=0.8in]{geometry}
\usepackage{amssymb}
\usepackage{hyperref}
\usepackage{ragged2e}
\usepackage{algorithm2e}
\usepackage[affil-it]{authblk}
\usepackage{abstract}

\usepackage[nodisplayskipstretch]{setspace} 
\begin{document} 

\title{\Large{\textbf{
Neutrino Mass Model in the Context of $\boldsymbol{\Delta(54) \otimes Z_2\otimes Z_3 \otimes Z_4}$ Flavor Symmetries with Inverse Seesaw Mechanism
 \vspace{-0.45em}}}}

\author{Hrishi Bora%
  \thanks{hrishi@tezu.ernet.in (Corresponding author)}}
  
 \author{Ng. K. Francis\thanks{francis@tezu.ernet.in}}
  \author{Animesh Barman\thanks{animesh@tezu.ernet.in}}
  \author{Bikash Thapa\thanks{bikash2@tezu.ernet.in}}

\affil{\vspace{-1.05em}Department of Physics, Tezpur University, Tezpur - 784028, India}

\date{\vspace{-5ex}}
\maketitle
%\linenumbers
\begin{center}
\textbf{\large{Abstract}}\\
\justify
Our analysis involves enhancing the $\Delta(54)$ flavor symmetry model with Inverse Seesaw mechanism along with two SM Higgs through the incorporation of distinct flavons. Additionally, we introduce supplementary $Z_2\otimes Z_3 \otimes Z_4$ symmetries to eliminate any undesirable components within our investigation.
The exact tri-bimaximal neutrino mixing pattern undergoes a deviation as a result of the incorporation of extra flavons, leading to the emergence of a non-zero reactor angle $\theta_{13}$ that aligns with the latest experimental findings. It was found that for our model the atmospheric oscillation parameter occupies the lower octant for normal hierarchy case. We also examine the parameter space of the model for normal hierarchy to explore the Dirac CP ($\delta_{CP}$),  Jarlskog invariant parameter ($J$) and the Neutrinoless double-beta decay parameter ($m_{\beta\beta}$) and found it in agreement with the neutrino latest data. Hence our model may be testable in the future neutrino experiments.
\end{center}

\hspace{-1.9em} PACS numbers: 12.60.-i, 14.60.Pq, 14.60.St
\newpage
\section{Introduction}
\label{sec:intro}

Despite the fact that the standard model (SM) has been verified by various particle physics experiments and observations, there are still several unresolved queations. These unanswered questions include, but are not limited to, the source of flavor structure, the matter-antimatter imbalance in the universe, the existence of additional neutrino flavors, and the enigmatic properties of dark matter and dark energy. The detection of neutrino oscillations in 1998 marked the first sign of physics beyond the standard model. However, many uncertainties still surround neutrino physics, such as whether the symmetry related to leptonic CP has been broken, what the lightest absolute neutrino mass value is, and whether the atmospheric mixing angle is maximal or not. Additionally, it remains unclear whether the neutrino masses follow the normal hierarchy ($m_1<m_2<m_3$) or the inverted hierarchy ($m_3 <m_1<m_2$). It is currently unknown whether neutrinos are Dirac particles or Majorana particles that violate lepton number. Neutrinoless double beta decay experiments, if observed, could confirm that neutrinos are indeed Majorana particles. The combined results of the KamLAND-Zen and GERDA experiments have established an upper limit for $m_{ee}$ within the range of 0.071–0.161 eV. Several recent reviews on neutrino physics can be found in references\cite{Nguyen:2018rlb, King:2014nza,King:2003jb,Cao:2020ans, Ahn:2014zja, mcdonald2016nobel, Nguyen:2020ehj, Hong:2022xjg, kajita2016nobel,de20212020,okada2021spontaneous, Ahn:2021ndu, PhongNguyen:2017meq,Buravov:2014dna, buravov2009elementary}.

The study of neutrino oscillation involves three mixing angles, two of which are large (the solar angle and the atmospheric angle) and one of which is relatively small (the reactor angle). In the context of tribimaximal mixing (TBM), the reactor angle is assumed to be zero and the CP phase cannot be determined or is not known. However, experiments such as the Daya Bay Reactor Neutrino Experiment ($sin^2 2\theta_{13}$ = $0.089 \pm 0.010 \pm 0.005$)\cite{DayaBay:2012fng} and RENO Experiment($sin^2 2\theta_{13}$ = $0.113 \pm 0.013 \pm 0.019$) \cite{RENO:2012mkc}have shown that the reactor mixing angle is not zero. Other experiments such as MINOS \cite{MINOS:2011amj}, Double Chooz\cite{DoubleChooz:2011ymz}, and T2K\cite{T2K:2011ypd} have also measured consistent nonzero values for the reactor mixing angle, rendering the TBM model unrealistic. As a result, alternative models or modifications must be considered to achieve realistic mixing\cite{RENO:2012mkc, DoubleChooz:2011ymz}.
The PMNS matrix, which is generally acknowledged, represents the blending between the various neutrino flavor states and their corresponding mass eigenstates. Within the context of a three-flavored model, the PMNS matrix is characterized by three mixing angles and three CP phases.

\begin{equation}
\label{eq:1}
    U_{PMNS}=
    \begin{pmatrix}
    c_{12} c_{13} & s_{12} c_{13} & s_{13}e^{-\iota\delta}\\
    -s_{12} c_{23}- c_{12} s_{23} s_{13} e^{\iota\delta} & c_{12} c_{23} - s_{12} s_{23} s_{13} e^{\iota\delta} & s_{23} c_{13}\\
    s_{12} s_{23} - c_{12} c_{23} s_{13} e^{\iota\delta} & -c_{12} s_{23} -s_{12} c_{23} s_{13} e^{\iota\delta} & \ c_{23} c_{13}
    \end{pmatrix}
    \cdot U_{Maj}
\end{equation}
where, $ c_{ij}=\cos{\theta_{ij}}$, $s_{ij}=\sin{\theta_{ij}}$.The diagonal matrix $U_{Maj}= diag (1, e^{\iota\alpha}, e^{\iota(\beta+\gamma)})$ holds the Majorana CP phases, $\alpha$ and $\beta$, which become detectable if neutrinos act as Majorana particles. Discovering neutrinoless double beta decay is likely the key to proving that neutrinos are Majorana particles, but these decays are yet to be observed. Symmetry is an essential factor in explaining this issue, as Wendell Furry proposed the Majorana nature of particles and studied a kinetic process similar to neutrinoless double beta decay\cite{Furry:1939qr, DellOro:2016tmg}. By producing a pair of electrons, this process breaks the lepton number by two units, generating Majorana neutrino masses as the electroweak symmetry breaks $(A,Z)\rightarrow (A,Z +2)+2e^- $. The large value of the lepton number violation scale($\Lambda \sim 10^{14}-10^{15} $GeV) is associated with the smallness of observed neutrino masses since they are zero in the standard model\cite{Bilenky:2012qi}. To generate non-zero neutrino mass, it is necessary to develop a new model beyond the standard model, such as effective theories that use the Weinberg operator\cite{Weinberg:1979sa}.

Several alternative frameworks to the standard model exist that can explain the origin of neutrino masses. These include the Seesaw Mechanism \cite{Yanagida:1980xy,Minkowski:1977sc, gell1979ramond, Mohapatra:1979ia,yanagida1979horizontal},  the Minimal Supersymmetric Standard Model (MSSM)\cite{Csaki:1996ks},Supersymmetry\cite{Ma:1998ias}, the Next-to-Minimal Supersymmetric Standard Model (NMSSM)\cite{Ellwanger:2009dp}, string theory\cite{Ibanez:2012zz}, Radiative Seesaw Mechanism\cite{Ma:2006km},  models based on extra dimensions\cite{Arkani-Hamed:1998wuz}, and various other models. Numerous neutrino experiments have demonstrated the presence of tiny but non-zero neutrino masses and provided evidence of flavor mixing\cite{aker2019improved,Francis:2014tea}.Several researchers have proposed different patterns of lepton mixing. Phenomenological patterns of neutrino mixing include Trimaximal (TM1/TM2)\cite{Albright:2008rp,He:2011gb,thapa2021resonant}, Tri-bimaximal (TBM)\cite{Harrison:2002er, Harrison:2002kp}, Bi-large mixing patterns\cite{Morisi:2009sc,Chen:2019egu,Ding:2019vvi} and Quasi-degenerate neutrino mass models. Additionally, several models based on non-abelian discrete symmetries\cite{King:2013eh} such as $A_4$\cite{Barman:2022hyq, Ishimori:2012fg,Ma:2006km, Vien:2014pta}, $S_3$\cite{Ma:2006km}, $S_4$, $\Delta(27)$\cite{Ma:2007wu,de2007neutrino,Harrison:2014jqa,CarcamoHernandez:2016piw} and $\Delta(54)$\cite{Ishimori:2008uc,Loualidi:2021qoq,Vien:2021fov} have been suggested to achieve tribimaximal mixing (TBM), and deviations from TBM are obtained by adding extra flavons. 

This study introduces a new approach, the inverse seesaw, using the $\Delta(54)$ flavour symmetry framework. The $\Delta$(54) symmetry can
manifest itself in heterotic string models on factorizable orbifolds, such as the $T^2/Z_3$ orbifold. In these string models, only singlets and triplets are observed as fundamental modes, while doublets are absent as fundamental modes. However, doublets have the potential to become fundamental
modes in magnetized/intersecting D-brane models. We can also suggest an extension to the Standard Model, utilizing $\Delta$(54) symmetry. We have the option to engage with both the singlets ($1_{1}, 1_{2}$) and doublets ($2_{1}, 2_{2}, 2_{3}, 2_{4} $)  representations of $\Delta(54)$, which allow us to represent quarks in different ways. This extension effectively incorporates the most recent experimental data for various properties within the quark sector, encompassing six quark masses, three quark mixing angles, and the CP-violating phase in this sector \cite{vien2021extension}.
 \captionsetup[table],
   {\begin{table}[h]
    \centering
    \scalebox{0.9}{
\begin{tabular}{c c c c c c c c c}
    \hline
       \textrm{Field}  &  $Q_{1L}$ & $ Q_{\alpha L}$ & $u_{\alpha R}$ & $u_{1R}$ & $d_{1R}$ & $d_{\alpha R}$ & $H$ & $\phi$ \\
     \hline
     \textrm{$\Delta(54)$}  &  $1_{+}$ &  $2_{2}$ & $1_{+}$ & $2_{2}$ & $1_{+}$ & $2_{2}$ & $1_{+}$ & $2_{2}$ \\
      \textrm{$U(1)$}  &  $1/6$ &  $1/6$ & $2/3$ & $2/3$ & $-1/3$ & $-1/3$ & $1/2$ & $1/2$ \\
     \hline 
\end{tabular} }
\end{table}}

A previous proposal by the authors of \cite{Borah:2017dmk} suggested Type I and inverse seesaw for Dirac neutrinos with A4 flavour symmetry. Lately, endeavors have been made in this particular area, particularly concerning the type I seesaw, type II seesaw  and type I+II for Dirac neutrinos within the $\Delta(54)$ symmetry \cite{kashav2022trimaximal}.The current work presents a more concise version of the inverse seesaw, utilizing two SM Higgs. To deviate from the specific TBM neutrino mixing pattern, we incorporated extra flavons  $\chi$, $\eta$, $\zeta$, $\xi$,$\Phi_{S}$ and $\phi$  under $\Delta(54)$ and expanded the flavon sector of the model. In addition, we incorporated a $Z_2 \otimes Z_3 \otimes Z_4$  symmetry into our model to prevent unwanted terms and facilitate the construction of specific coupling matrices. We modified the structure of the neutrino mass matrix $M_{\nu}$ that characterizes the masses of neutrinos  and conducted symmetry-based studies. This approach allows us to thoroughly examine $M_{\nu}$, and investigate the Jarlskog invariant parameter ($J$) and the NDBD parameter ($m_{ee}$). The approach we take sets our work apart from that of others.

Our paper's structure is arranged as follows: Section 2 presents an outline of the model's framework, including the fields and their symmetrical transformation properties. Section 3 contains a numerical analysis and examination of the neutrino phenomenology results. Lastly, in section 4, we offer our concluding remarks.

\begin{table}[t]
\centering
  \begin{tabular}{ | l | c | r |}
    \hline
    Parameters & NH (3$\sigma$) & IH (3$\sigma$) \\ \hline
    $\Delta{m}^{2}_{21}[{10}^{-5}eV^{2}]$ & $6.82 \rightarrow 8.03$ & $6.82 \rightarrow 8.03$ \\ \hline
    $\Delta{m}^{2}_{31}[{10}^{-3}eV^{2}]$ & $2.428 \rightarrow 2.597$ & $-2.581 \rightarrow -2.408 $\\ \hline
    $\sin^{2}\theta_{12}$ & $0.270 \rightarrow 0.341$ & $0.270 \rightarrow 0.341$ \\ \hline
     $\sin^{2}\theta_{13}$ & $0.02029 \rightarrow 0.02391$ & $0.02047 \rightarrow 0.02396$ \\ \hline
    $\sin^{2}\theta_{23}$ & $0.406 \rightarrow 0.620$ & $0.410 \rightarrow 0.623$ \\ \hline
    $\delta_{CP}$ & $108 \rightarrow 404$ & $192 \rightarrow 360$ \\ \hline
  \end{tabular}
    \caption{The $3\sigma$ range of neutrino oscillation parameters from NuFIT 5.2 (2022) \cite{Esteban:2020cvm}}
    \label{tab:1}
\end{table}

\section{Framework of the Model}
To achieve the realization of the Inverse seesaw mechanism, it is necessary to expand the fermion sector of the Standard Model. We introduced Vector like (VL) fermions N and S which are all Standard Model gauge singlets. The left and right-handed fields are distinguished with a subscript 1 and 2 respectively.  $\nu_R$ is a right handed neutrino. It gives us a Dirac mass matrix in the Lagrangian. The $\Delta(54)$  group includes irreducible representations $1_+$, $1_-$, $2_1$, $2_2$, $2_3$, $2_4$, $3_{1(1)}$, $3_{1(2)}$, $3_{2(1)}$ and $3_{2(2)}$. While there are four triplets, the products of $3_{1(1)}\otimes3_{1(2)}$ , $3_{2(1)}\otimes3_{2(2)}$ ,$3_{1(1)}\otimes3_{2(2)}$ and $3_{1(2)}\otimes3_{2(1)}$ lead to the trivial singlet.

The multiplication rules are as follows:

$$3_{1(1)}\otimes3_{1(1)}=3_{1(2)} \oplus 3_{1(2)} \oplus3_{2(2)}$$
$$3_{1(2)}\otimes3_{1(2)}=3_{1(1)}\oplus 3_{1(1)}\oplus 3_{2(1)}$$
$$3_{2(1)}\otimes3_{2(1)}=3_{1(2)}\oplus 3_{1(2)}\oplus 3_{2(2)}$$
$$3_{2(2)}\otimes3_{2(2)}=3_{1(1)}\oplus 3_{1(1)}\oplus 3_{2(1)}$$
$$3_{1(1)}\otimes3_{1(2)}=1_{1}\oplus 2_{1}\oplus 2_{2}\oplus 2_{3}\oplus 2_{4}$$
$$3_{1(2)}\otimes3_{2(1)}=1_{2}\oplus 2_{1}\oplus 2_{2}\oplus 2_{3}\oplus 2_{4}$$
$$3_{2(1)}\otimes3_{2(2)}=1_{1}\oplus 2_{1}\oplus 2_{2}\oplus 2_{3}\oplus 2_{4}$$
$$3_{1(1)}\otimes3_{2(2)}=1_{2}\oplus 2_{1}\oplus 2_{2}\oplus 2_{3}\oplus 2_{4}$$

For two triplets
$$a = (a_1,a_2,a_3)$$
$$b = (b_1,b_2,b_3)$$
We can write
$$1_{1} = (ab)_{+} = a_1 b_1 +a_2 b_2 +a_3 b_3$$
$$1_{2} = (ab)_{-} = a_1 b_1 +a_2 b_2 +a_3 b_3$$

\begin{table}[t]
    \centering
    \scalebox{0.8}{
    \begin{tabular}{c c c c c c c c c c c c c c c c}
    \hline
       \textrm{Field}  &  L & $ l= 
    e^{c},\mu^{c},\tau^{c}$ & $H_1$ & $H_2$ & $\nu_{R}$ & $N_{1}$ & $N_{2}$ & $S_1$ & $S_2$ & $\chi$ & $\eta$ & $\zeta$ & $\xi$ & $\Phi_{S}$ & $\phi$ \\
     \hline
     \textrm{$\Delta(54)$}  &  $3_{1(1)}$ &  $3_{1(1)}$ & $1_{2}$ & $1_{1}$ & $3_{2(2)}$ & $3_{1(1)}$ & $3_{2(2)}$ & $3_{1(1)}$ & $3_{2(2)}$ & $3_{2(1)}$ &  $3_{2(2)}$ & $1_{2}$ & $3_{2(1)}$ & $3_{2(1)}$ & $1_{2}$\\
     \textrm{Z}$_2$  &  1 & 1 & 1 & 1 & -1 & -1 & 1 & -1 & 1 & 1 & 1 & -1 & -1 & -1 & -1\\
    \textrm{Z}$_3$  &  $\omega$ & $\omega$ & 1 & 1 & 1 & 1 & 1 & 1 & 1 & 1 & 1 & 1 & $\omega$ & $\omega$ & 1\\
\textrm{Z}$_4$  &  1 & 1 & 1 & 1 & -i & 1 & -i & -i & 1 & 1 & 1 & 1 & 1 & 1 & -i\\
\textrm{U(1)}  &  1 & 1 & 0 & 0 & 1 & 1 & 1 & 1 & 1 & 0 & 0 & 0 & 0 & 0 & 0\\
     \hline
    \end{tabular}}
    \caption{Full particle content of our model}
    \label{tab:2}
\end{table}

Our model is based on the $\Delta54$ model where we introduced additional flavons $\chi$, $\xi$, $\eta$ ,$\zeta$ , $\Phi_{S}$ and $\phi$ to get the deviation from the precise TBM pattern of neutrino mixing\cite{2010}. We put extra symmetry $Z_2\otimes Z_3 \otimes Z_4$ to avoid unwanted terms. Table \ref{tab:2} provides details regarding the composition of the particles and corresponding charge assignment in accordance with the symmetry group. The triplet representation of $\Delta54$ is used to assign the right-handed charged leptons and the left-handed lepton doublets.

The invariant Yukawa Lagrangian is as follows \footnote{Considering terms upto dimension-5.}:

\begin{align}
  \mathcal{L} = & \sum_{k=e,\mu,\tau} \frac{y^{ij}_k}{\Lambda}(\chi_{i} l_{j} )\Bar{L_{k}}H_1  + \frac{\Bar{L} \Tilde{H_1} N_{1}}{\Lambda}y_{\xi} \xi + \frac{\Bar{L} \Tilde{H_1} N_{1}}{\Lambda} y_{s} \Phi_{s} + \frac{\Bar{L} \Tilde{H_2} N_{1}}{\Lambda} y_{a} \Phi_{s}   \\                 
 & + \frac{y_{\scaleto{RN}{3pt}}}{\Lambda} \Bar{\nu_{R}} N_{2} \eta \zeta + y_{\scaleto{NS}{3pt}} \Bar{S_{1}} N_{2} \zeta + y_{\scaleto{NS}{3pt}}^{\prime} \Bar{S_{2}} N_{1} \zeta + \frac{y_{s}}{\Lambda^2}\Bar{S_{1}} S_{2} \phi 
\end{align}

We consider the vacuum expectation values as,
\begin{align*}
\langle \Phi_S \rangle& =(v_s ,v_s,v_s)&
\langle \chi \rangle& =(v_{\chi},v_{\chi},v_{\chi})&
\langle \xi \rangle& =(v_{\xi}, v_{\xi}, v_{\xi})&\\
\langle \eta \rangle& =(v_{\eta},v_{\eta},v_{\eta})&
\langle \zeta \rangle& =(v_{\zeta})&
\langle \phi \rangle& =(v_{\phi}) &  
\end{align*}

The charged lepton mass matrix is given as 
\begin{equation}
    M_l= \frac{v_{\scaleto{\chi}{3pt}}H_{1}}{\Lambda}
    \begin{pmatrix}
    y_e & 0 & 0\\
    0 & y_\mu  & 0\\
    0 & 0 &  y_\tau 
    \end{pmatrix}
\end{equation}

where, $y_e$, $y_{\mu}$ and $y_{\tau}$ are the coupling constants.

\subsection{Effective neutrino mass matrix}

By utilizing the Lagrangian presented, it is possible to derive the mass matrices that pertain to the neutrino sector once both $\Delta{54}$ and electroweak symmetry breaking have occurred. The essence of the ISS theory lies in the assurance that the neutrino masses remain small by postulating a small $M_{S}$ scale. To reduce the right-handed neutrino masses to the TeV scale, it is necessary for the $M_{S}$ scale to be at the KeV level. The inverse seesaw model is a TeV-scale seesaw model that allows heavy neutrinos to stay as light as a TeV while permitting Dirac masses to be as substantial as those of charged leptons, all while maintaining compatibility with light neutrino masses in the sub-eV range.

  \begin{equation}
     M_{RN}=  \frac{y_{\scaleto{RN}{3pt}}}{\Lambda}
    \begin{pmatrix}
    v_\eta v_\zeta & 0 & 0\\
    0 &  v_\eta v_\zeta & 0\\
    0 & 0 &   v_\eta v_\zeta
    \end{pmatrix}
\end{equation}

\begin{equation}
M_{NS}=   y_{\scaleto{NS}{3pt}} \begin{pmatrix}
     v_{\zeta} & 0 & 0\\
    0 &  v_{\zeta} & 0\\
    0 & 0 &  v_{\zeta}
    \end{pmatrix}
\end{equation}

\begin{equation}
    M_{NS}^{\prime}=  y_{\scaleto{NS}{3pt}}^{\prime}
 \begin{pmatrix}
    v_{\zeta} & 0 & 0\\
    0 & v_{\zeta} & 0\\
    0 & 0 &  v_{\zeta}
    \end{pmatrix}
    \end{equation}

\begin{equation}    
M_{S}=  \frac{y_s}{\Lambda^{2}}\begin{pmatrix}
     v_{\phi} & 0 & 0\\
    0 &  v_{\phi} & 0\\
    0 & 0 &   v_{\phi}
    \end{pmatrix} 
\end{equation}

 \begin{equation}
 M_{\nu N}= \frac{v}{\Lambda}        \begin{pmatrix} 
    y_{\scaleto{\xi}{6pt}} v_{\scaleto{\xi}{6pt}} & y_{s}v_{s} + y_{a} v_a & y_{s}v_{s} - y_{a} v_a\\
   y_{s}v_{s} - y_{a} v_a &   y_{\scaleto{\xi}{6pt}} v_{\scaleto{\xi}{6pt}} & y_{s}v_{s} + y_{a} v_a\\
    y_{s}v_{s} + y_{a} v_a & y_{s}v_{s} - y_{a} v_a &   y_{\scaleto{\xi}{6pt}} v_{\scaleto{\xi}{6pt}}
    \end{pmatrix}
   \end{equation}

Effective neutrino mass matrix is given by
\begin{equation} 
m_\nu = M_{RN}(M_{NS}^{\prime})^{-1}M_{S}M_{NS}^{-1}M_{\nu N}
\end{equation}
\begin{equation}
   = \lambda
    \begin{pmatrix}
     x & s+a & s-a\\
    s-a   & x &  s+a\\
    s+a & s-a &  x
    \end{pmatrix}   
\end{equation}

\begin{equation}
   = 
    \begin{pmatrix}
     x^{\prime} & s^{\prime}+a^{\prime} & s^{\prime}-a^{\prime}\\
    s^{\prime}-a^{\prime}   & x^{\prime} &  s^{\prime}+a^{\prime}\\
    s^{\prime}+a^{\prime} & s^{\prime}-a^{\prime} &  x^{\prime}
    \end{pmatrix}   
\end{equation}

where, $x = y_{\scaleto{\xi}{6pt}}v_{\scaleto{\xi}{6pt}}$ , $s= y_s v_s$ and $a = y_a v_a$ \\
 We now define the Hermitian matrix as
 \begin{equation} 
M_\nu = m_\nu m_{\nu}^\dag
\end{equation} 

\begin{equation} 
  = 
    \begin{pmatrix}
     x^{\prime 2} + 2(s^{\prime2}+a^{\prime2})   &   s^{\prime2}-a^{\prime2} + 2s^{\prime}x^{\prime}   &      s^{\prime2} -a^{\prime2} + 2s^{\prime}x^{\prime}\\
     s^{\prime2}-a^{\prime2} + 2s^{\prime}x^{\prime}    &   x^{\prime2} + 2(s^{\prime2}+a^{\prime2})  &       s^{\prime2} -a^{\prime2}  +2s^{\prime}x^{\prime}  \\
     s^{\prime2} -a^{\prime2} + 2s^{\prime}x^{\prime} &   s^{\prime2}-a^{\prime2} +  2s^{\prime}x^{\prime}   &     x^{\prime2} + 2(s^{\prime2}+a^{\prime2})
    \end{pmatrix}
\end{equation} 

where, $\lambda = \frac{v v_\eta v_{\phi} y_{\scaleto{RN}{3pt}} y_{s}} {{\Lambda}^4 v_{\zeta} y_{\scaleto{NS}{3pt}} y_{\scaleto{NS}{3pt}}^{\prime}} $ , $ x^{\prime} = x{\lambda}$,  $ s^{\prime} = s{\lambda}$ and $ a^{\prime} = a{\lambda}$

\section{Numerical Analysis and results} 

In the earlier section, we illustrated the method of enhancing the $\Delta54$ model by incorporating supplementary flavons. The subsequent section comprises a quantitative investigation aimed at examining the effectiveness of the parameters in generating a departure from TBM in neutrino mixing. The outcomes of this analysis, restricted to the normal hierarchy scenario, will be deliberated.

\noindent The neutrino mass matrix $m_\nu$ can be diagonalized by the PMNS matrix $U$ as
\begin{equation}
    \label{eq:12}
    U^\dagger m^{(i)}_\nu U^* = \textrm{diag(}m_1, m_2, m_3 \textrm{)}
\end{equation}
 We can numerically calculate $U$ using the relation $U^\dagger M_\nu U = \textrm{diag(}m_1^2, m_2^2, m_3^2 \textrm{)}$, where $M_\nu = m_\nu m^{\dagger}_\nu$. The neutrino oscillation parameters $\theta_{12}$, $\theta_{13}$, $\theta_{23}$ and $\delta$ can be obtained from $U$ as
\begin{equation}
    \label{eq:13}
    s_{12}^2 = \frac{\lvert U_{12}\rvert ^2}{1 - \lvert U_{13}\rvert ^2}, ~~~~~~ s_{13}^2 = \lvert U_{13}\rvert ^2, ~~~~~~ s_{23}^2 = \frac{\lvert U_{23}\rvert ^2}{1 - \lvert U_{13}\rvert ^2}
\end{equation}

and $\delta$ may be given by
\begin{equation}
    \label{eq:14}
    \delta = \textrm{sin}^{-1}\left(\frac{8 \, \textrm{Im(}h_{12}h_{23}h_{31}\textrm{)}}{P}\right)
\end{equation}
with 
\begin{equation}
    \label{eq:15}
     P = (m_2^2-m_1^2)(m_3^2-m_2^2)(m_3^2-m_1^2)\sin 2\theta_{12} \sin 2\theta_{23} \sin 2\theta_{13} \cos \theta_{13}
\end{equation}

To assess how the neutrino mixing parameters compare to the most recent experimental data \cite{Esteban:2020cvm}, we adjusted the modified $\Delta54$ model to fit the experimental data by minimizing the subsequent $\chi^2$ function:

\begin{equation}
	\label{eq:16}
	\chi^2 = \sum_{i}\left(\frac{\lambda_i^{model} - \lambda_i^{expt}}{\Delta \lambda_i}\right)^2,
\end{equation}

where $\lambda_i^{model}$ is the $i^{th}$ observable predicted by the model, $\lambda_i^{expt}$ stands for  $i^{th}$ experimental best-fit value and $\Delta \lambda_i$ is the 1$\sigma$ range of the observable.

\begin{figure}[t]
     \centering
     \begin{subfigure}[b]{0.46\textwidth}
         \centering
         \includegraphics[width=\textwidth]{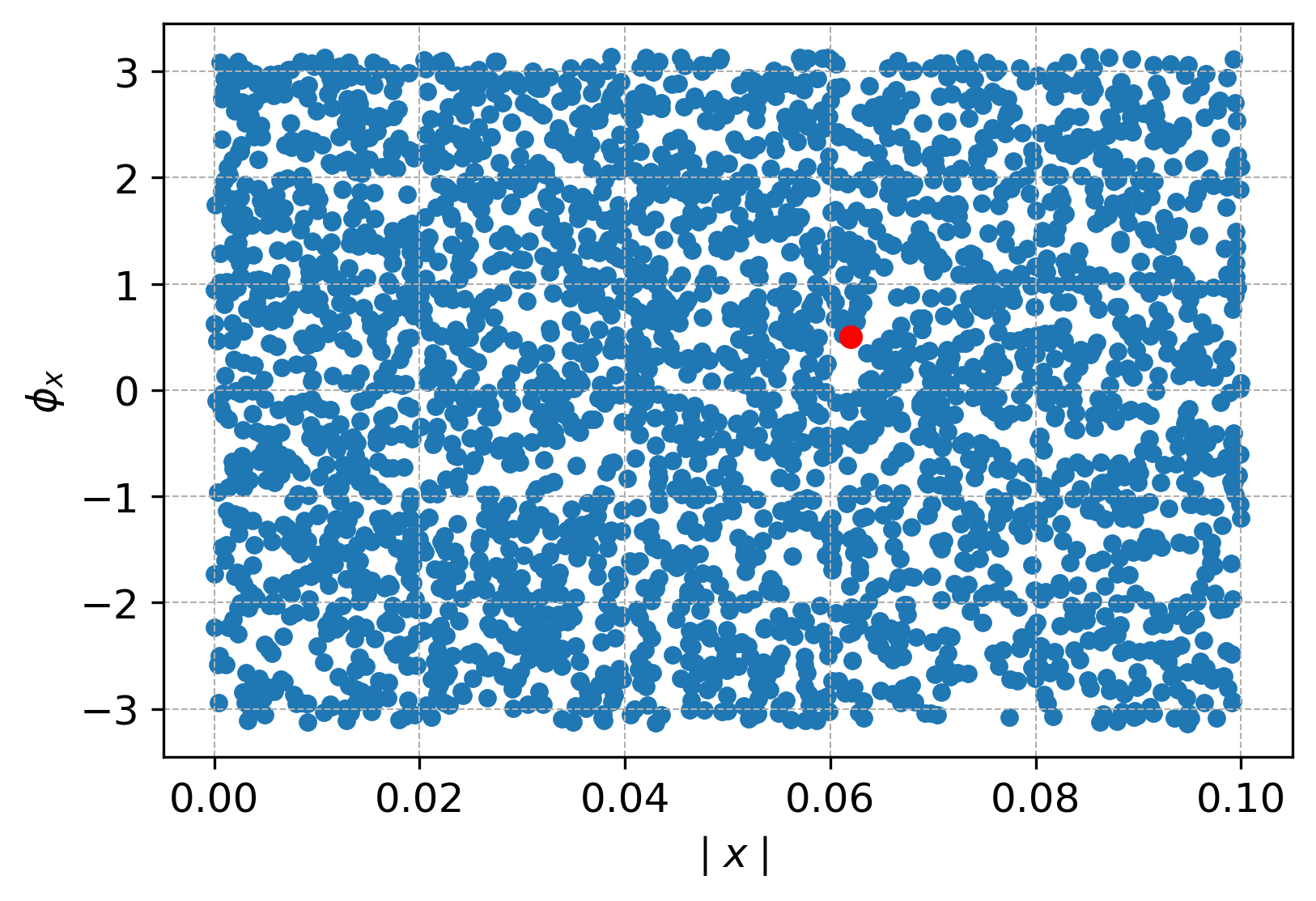}
     \end{subfigure}
     \hfill
     \begin{subfigure}[b]{0.46\textwidth}
         \centering
         \includegraphics[width=\textwidth]{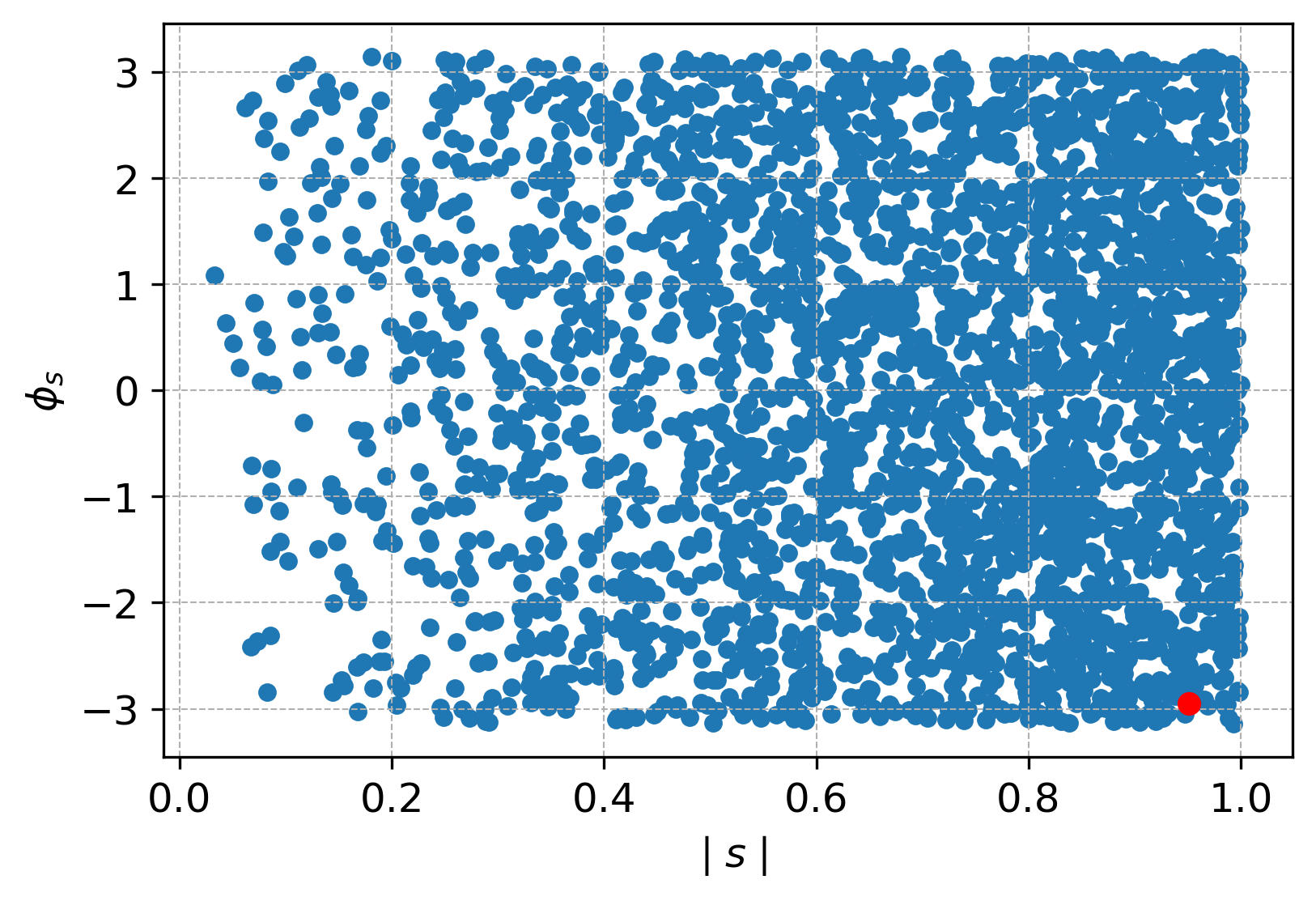}
     \end{subfigure}

     \vspace{1em}
     \begin{subfigure}[b]{0.46\textwidth}
         \centering
         \includegraphics[width=\textwidth]{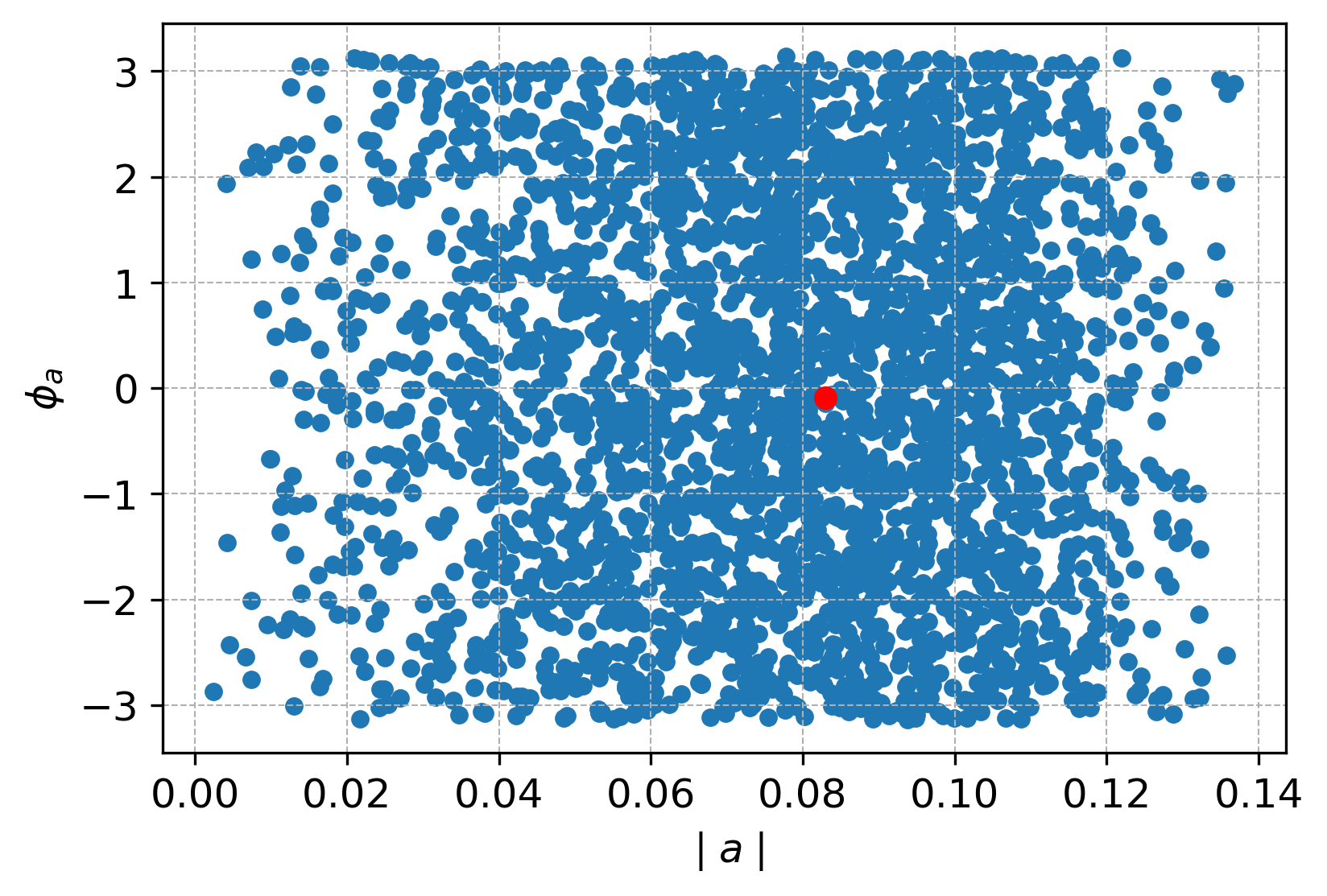}
     \end{subfigure}
  
    \caption{Allowed regions of the model parameters $\lvert x\rvert$, $\phi_x$, $\lvert s\rvert$, $\phi_s$ and $\lvert a\rvert$, $\phi_a$ in NH. The best fit values are indicated by red dots.}
    \label{fig:1}
\end{figure}

The parameter of the model space is illustrated in Fig.\ref{fig:1}, with restrictions based on the 3$\sigma$ limit of neutrino oscillation data presented in Table\ref{tab:1}. The illustration indicates a strong interdependence among various parameters of the model. The best-fit values for $\lvert x\rvert$, $\lvert s\rvert$, $\lvert a\rvert$, $\phi_x$, $\phi_s$ and $\phi_a$ obtained are (0.062, 0.951, 0.083, 0.505$\pi$, -2.953$\pi$, -0.092$\pi$).

\begin{figure}[t]
     \centering
     \begin{subfigure}[b]{0.46\textwidth}
         \centering
         \includegraphics[width=\textwidth]{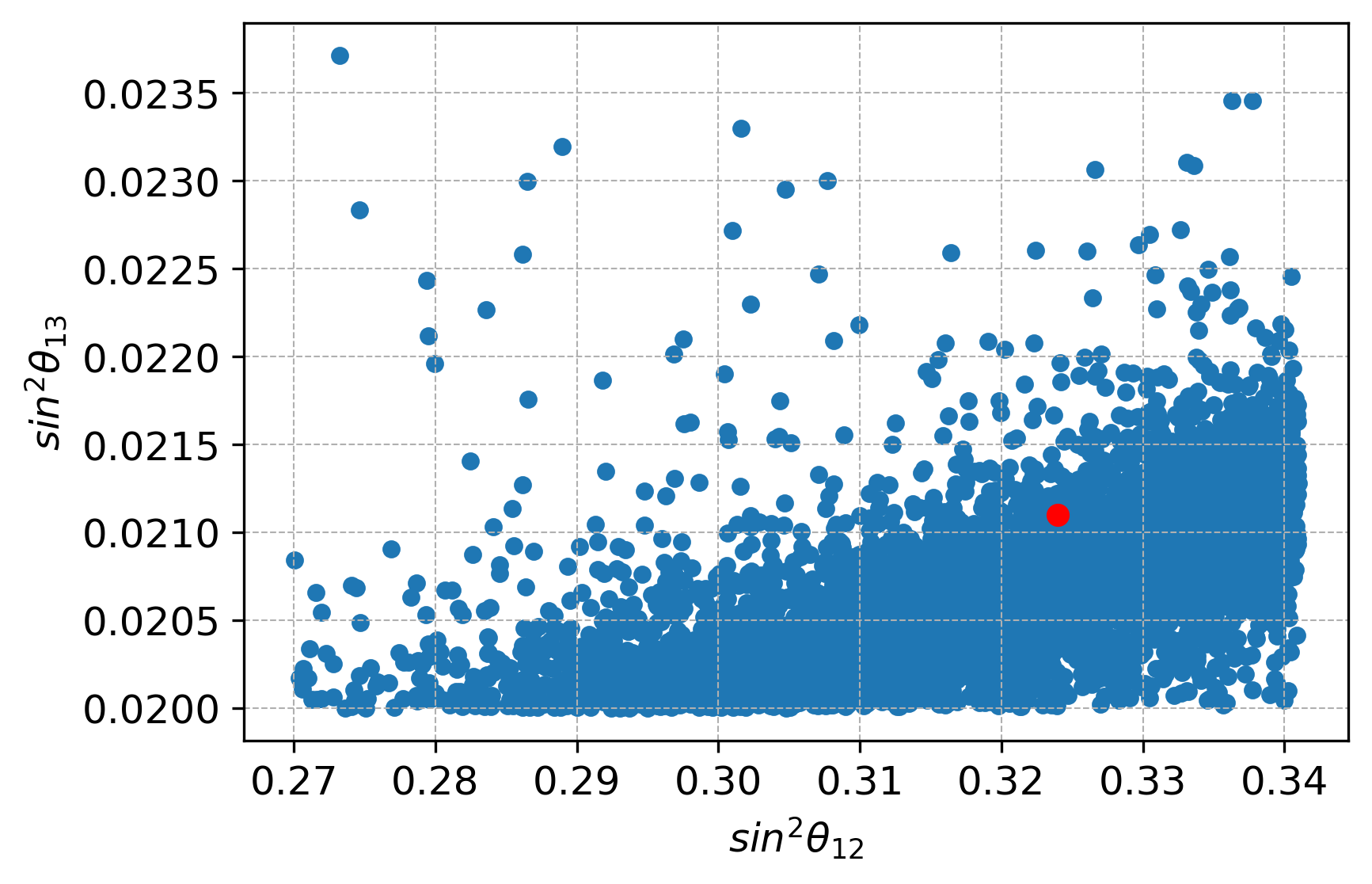}
     \end{subfigure}
     \hfill
     \begin{subfigure}[b]{0.46\textwidth}
         \centering
         \includegraphics[width=\textwidth]{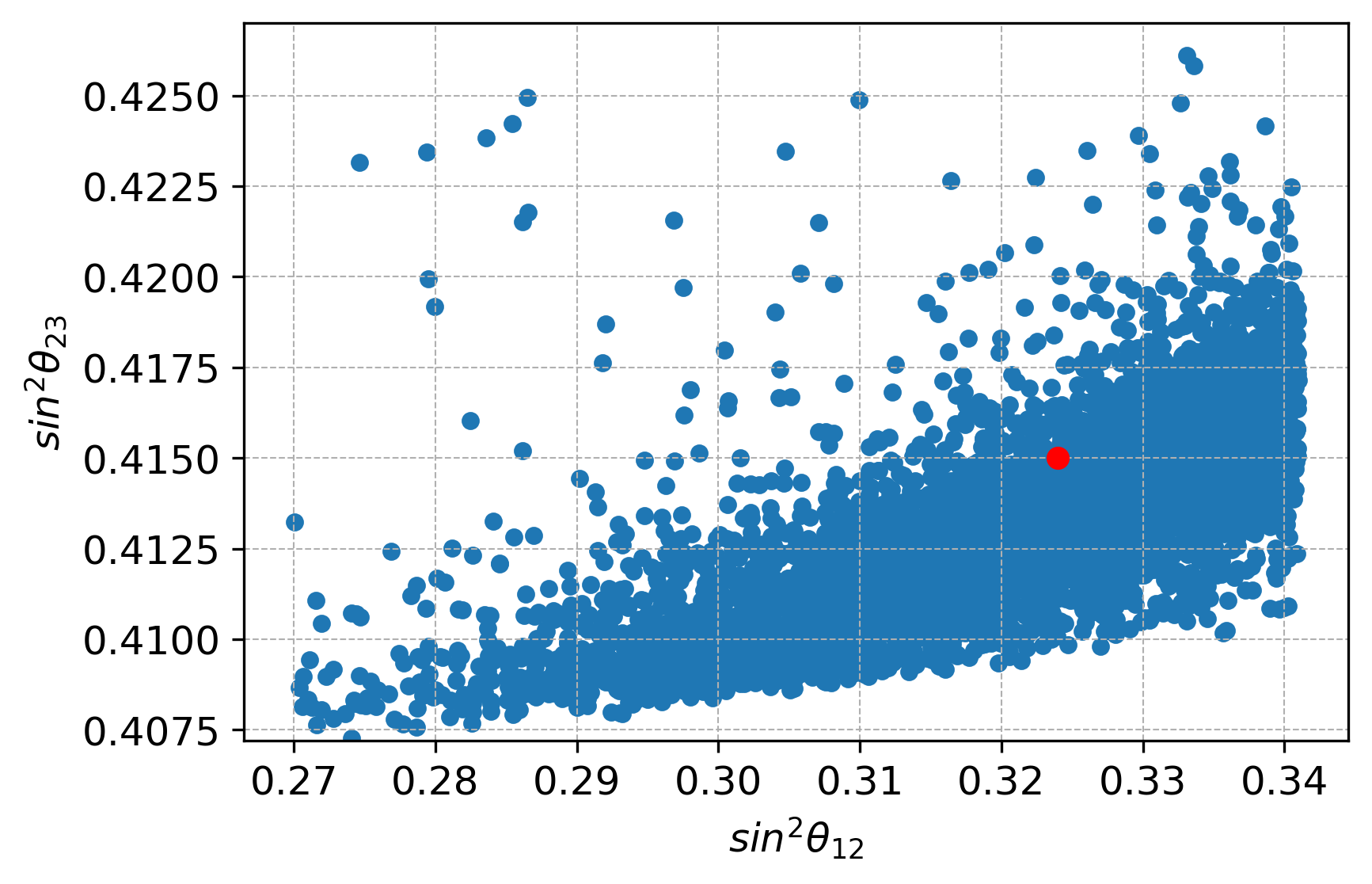}
     \end{subfigure}

     \vspace{1em}
     \begin{subfigure}[b]{0.46\textwidth}
         \centering
         \includegraphics[width=\textwidth]{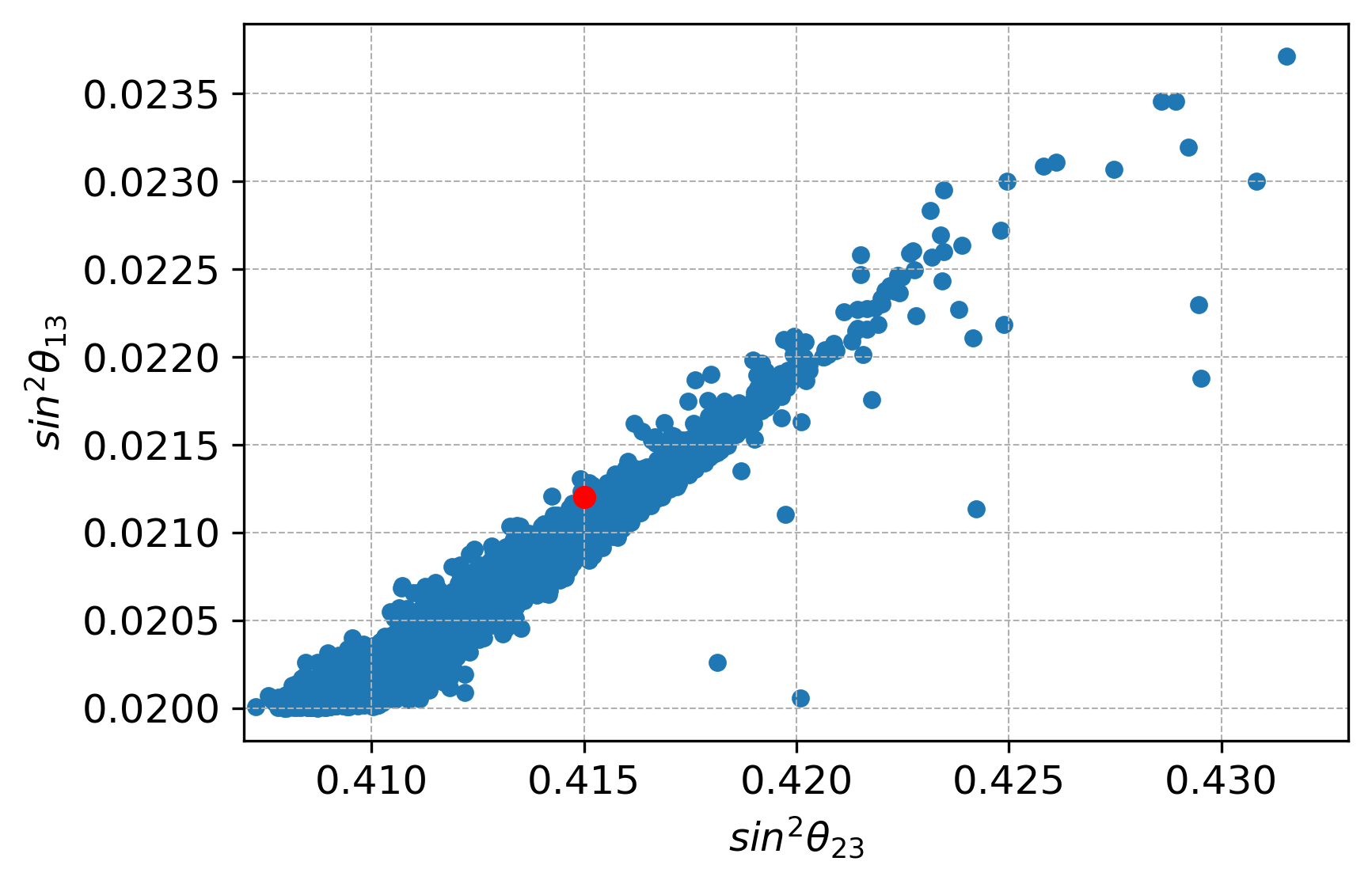}
     \end{subfigure}
     \hfill
     \begin{subfigure}[b]{0.46\textwidth}
         \centering
         \includegraphics[width=\textwidth]{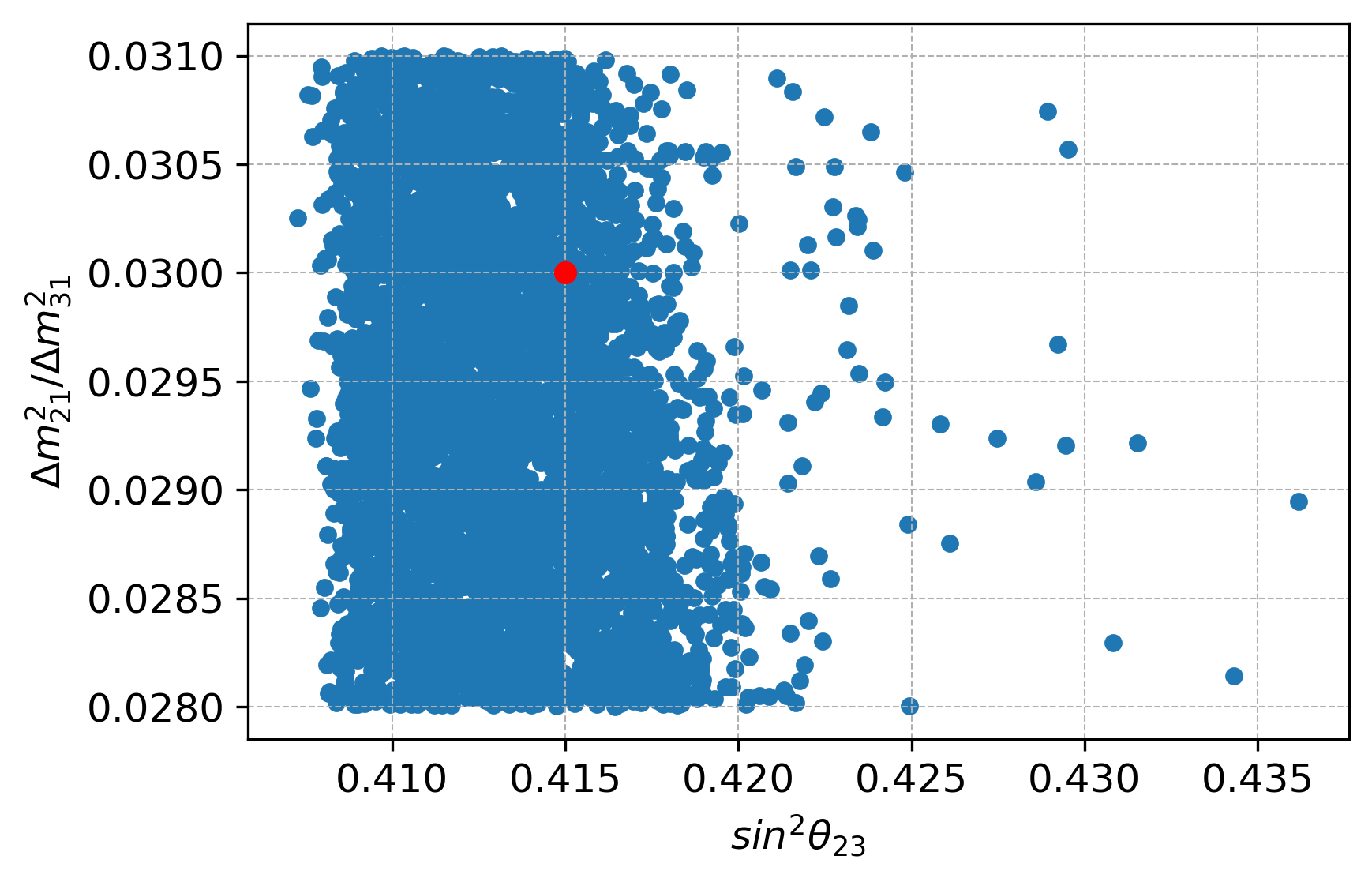}
     \end{subfigure}
    \caption{Correlation among the mixing angles and among mixing angle and ratio of mass-squared difference. The best fit value is indicated by the red dot.}
    \label{fig:2}
\end{figure}

Fig. \ref{fig:2} predicts the expected values of the neutrino oscillation parameters within the model for NH. The best fit values of $\sin^2 \theta_{12}$, $\sin^2 \theta_{13}$ and $\sin^2 \theta_{23}$ are (0.320, 0.0214, 0.417) which are within the 3$\sigma$ range of experimental measurements. Additional parameters such as $\Delta m_{21}^2$, $\Delta m_{31}^2$ and $\delta_{CP}$ have their best-fit values, corresponding to $\chi^2$-minimum, at ($6.88 \times 10^{-5}$ eV, $2.50 \times 10^{-3}$ eV, $0.115 \pi$). 

\begin{figure}[t]
     \centering

     \begin{subfigure}[b]{0.46\textwidth}
         \centering
         \includegraphics[width=\textwidth]{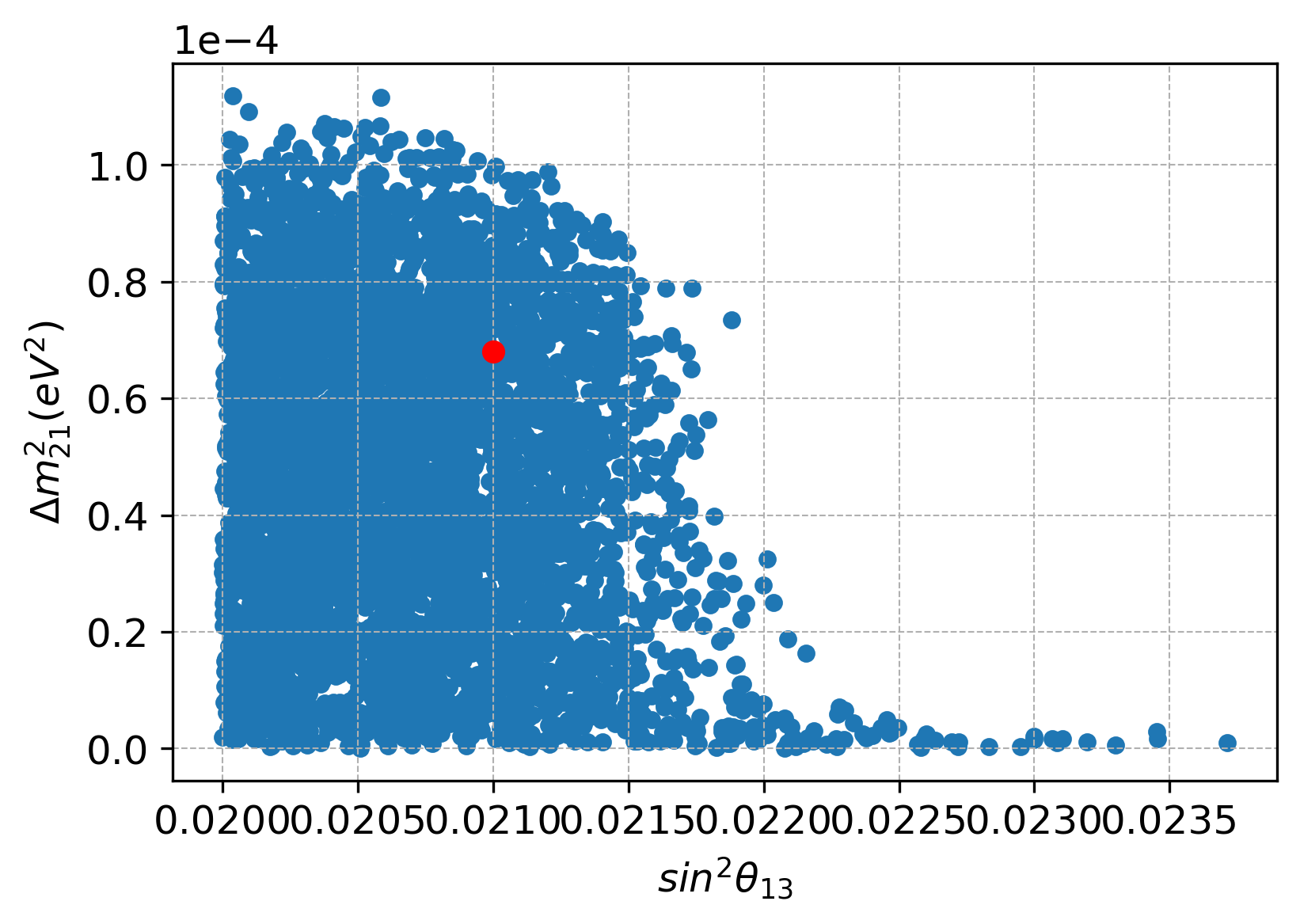}
     \end{subfigure}
     \hfill
     \begin{subfigure}[b]{0.46\textwidth}
         \centering
         \includegraphics[width=\textwidth]{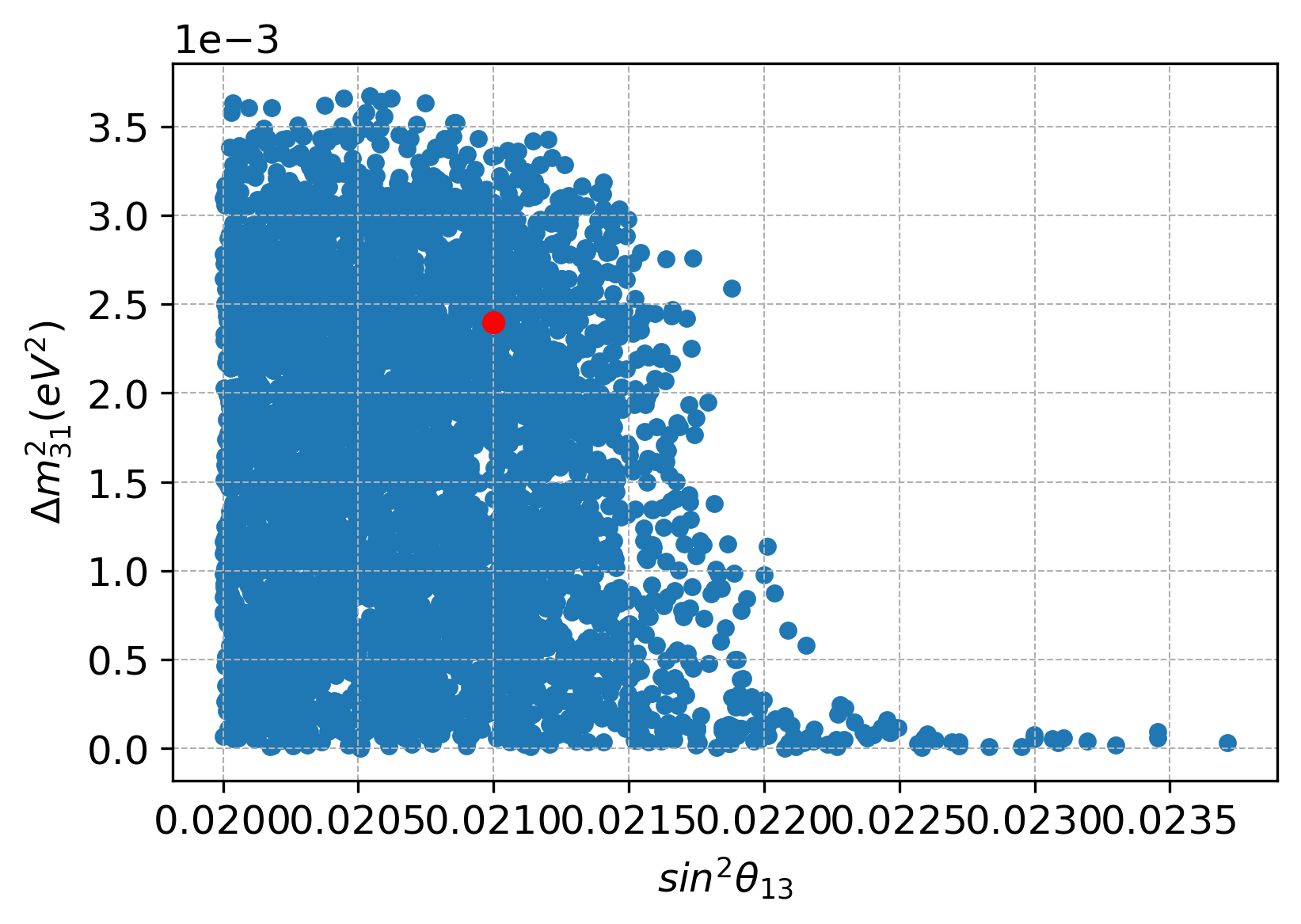}
     \end{subfigure}

       \vspace{1em}

     \begin{subfigure}[b]{0.46\textwidth}
         \centering
         \includegraphics[width=\textwidth]{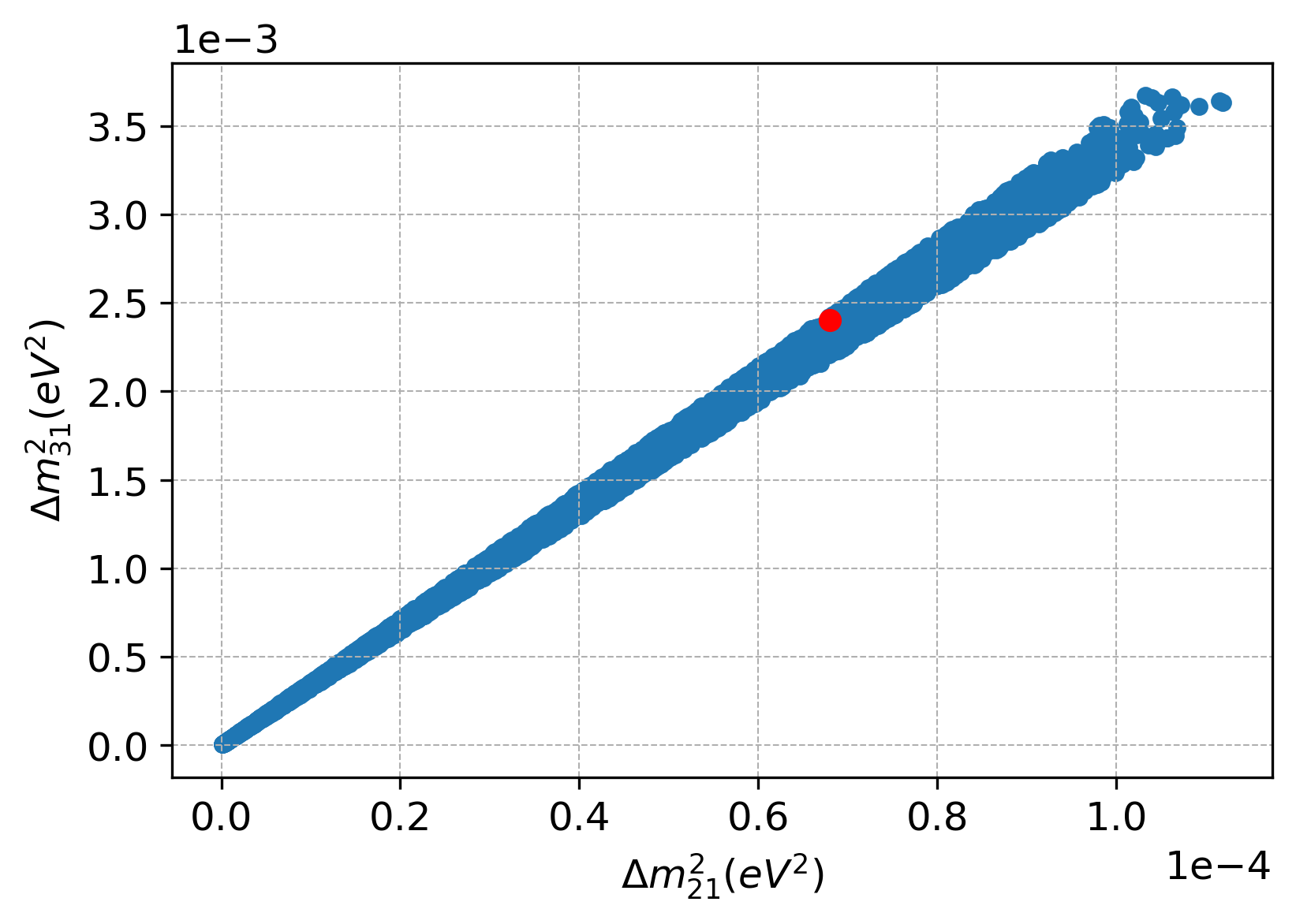}
     \end{subfigure}
    \caption{Variation of the mass squared differences with mixing angles and correlation between two mass squared differences.}
    \label{fig:3}
\end{figure}

Fig. \ref{fig:3} gives the correlation between the two mass square difference predicted by the model for NH. The $\Delta m_{21}^2$ and $\Delta m_{31}^2$  have their best-fit values, corresponding to $\chi^2$-minimum, at ($6.88 \times 10^{-5}$ eV, $2.507 \times 10^{-3}$ eV) respectively.

\begin{figure}[t]
     \centering
     \begin{subfigure}[b]{0.46\textwidth}
         \centering
         \includegraphics[width=\textwidth]{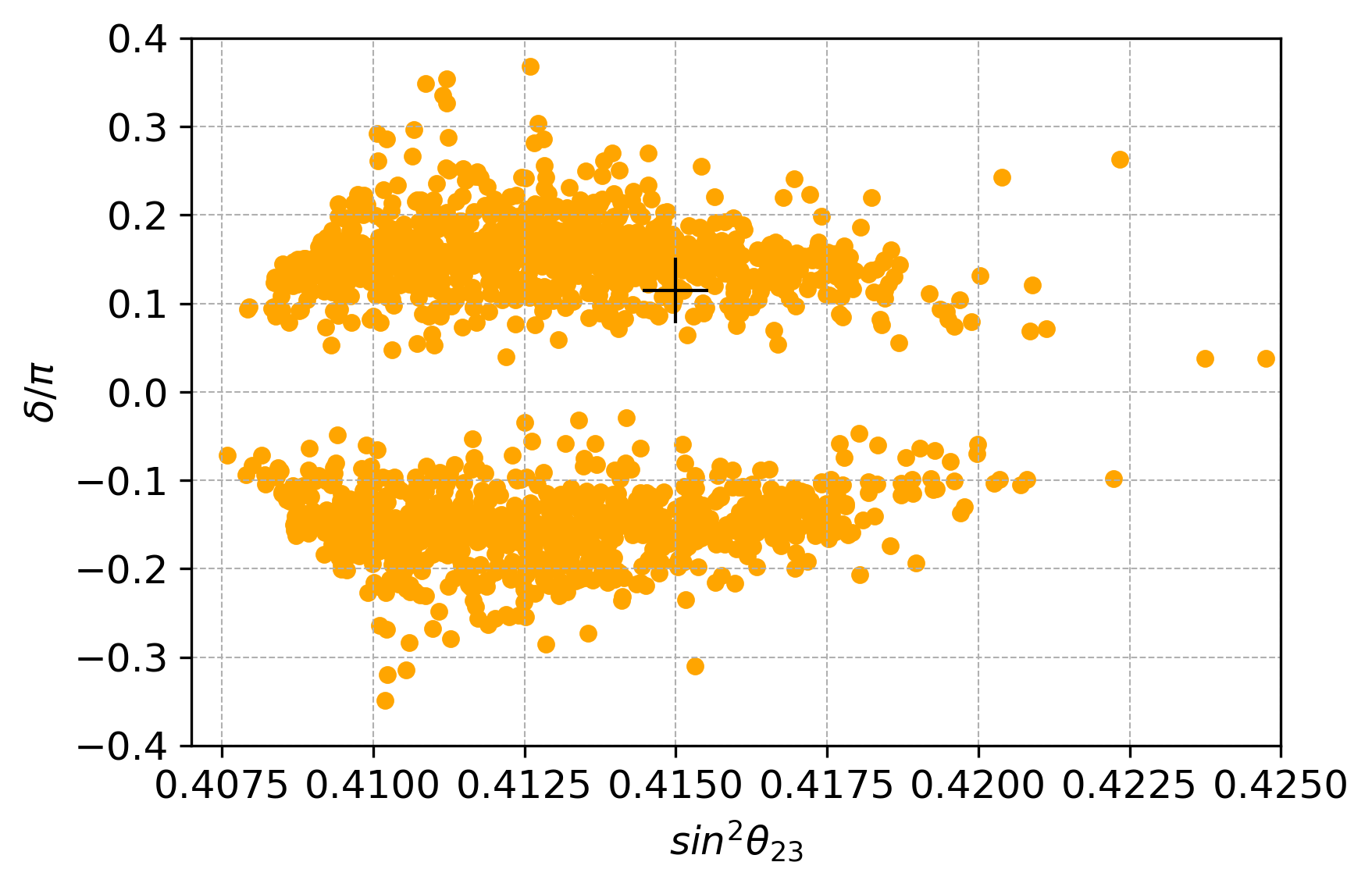}
     \end{subfigure}
     \hfill
     \begin{subfigure}[b]{0.46\textwidth}
         \centering
         \includegraphics[width=\textwidth]{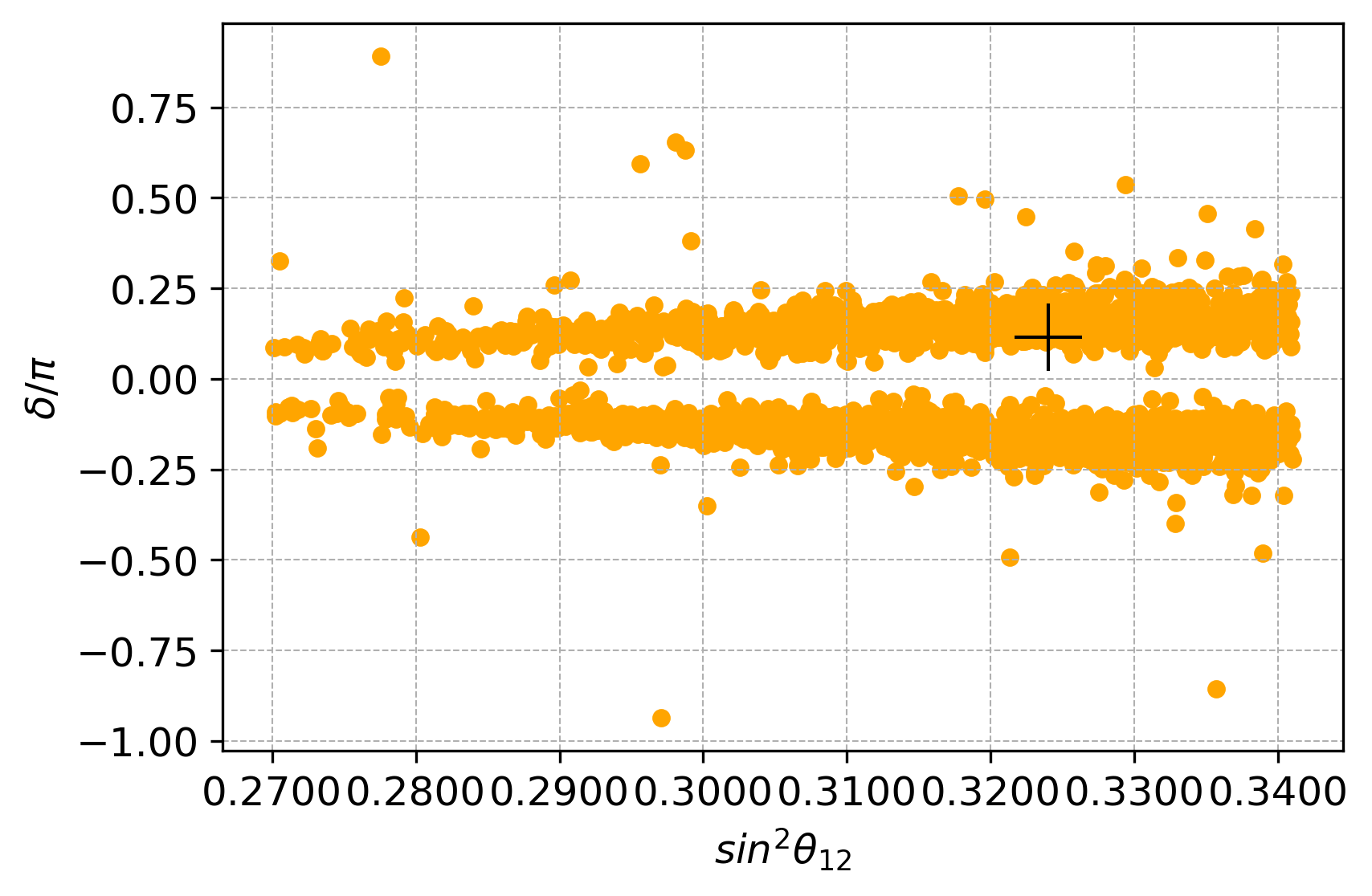}
     \end{subfigure}

    \caption{Correlation between Dirac CP ($\delta_{CP}$) and atmospheric mixing angle($sin^2\theta_{23}$) and solar mixing angle($sin^2\theta_{12}$) respectively. The best fit value for 3$\sigma$ range is given by red dot.}
    \label{fig:4}
\end{figure}

Fig. \ref{fig:4} gives the correlation between the Dirac CP phase and atmospheric angle and also Dirac CP with the ratio of the two mass squared difference. The best fit value of $\delta_{CP}$ is predicted to be around 0.115$\pi$.

Thus, the model defined in this work indicates clear deviation from tri-bimaximal mixing.

It is clear from the correlation plot among the neutrino oscillation parameters that the mixing of neutrinos deviates from the TBM mixing in NH. The prediction of mixing angle $\theta_{23}$ in NH scenario indicates a preference towards the lower octant. With the modification of $\Delta54$ model with additional term, it is possible to deviate from TBM mixing. The Table \ref{tab:3} shows the best-fit values of the various oscillation parameters.

\begin{table}[!ht]
    \centering
    \begin{tabular}{c c }
    \hline
       \textrm{Parameter}  &  \textrm{NH} \\
     \hline
     
     $\sin^2\theta_{12}$ & 0.320  \\
     
     $\sin^2\theta_{13}$ & 0.0214 \\
     
     $\sin^2\theta_{23}$ & 0.417 \\
     
     $\delta_{CP}/\pi$ & 0.115   \\
     
     $\Delta m_{21}^2$ & $6.880 \times 10^{-5}$ \\
     
     $\Delta m_{31}^2$ & $2.507 \times 10^{-3}$  \\

    \hline
    \end{tabular}
    \caption{Best-fit values for different parameters predicted by the model}
    \label{tab:3}
\end{table}

\begin{figure}[!ht]
     \centering
     \begin{subfigure}[b]{0.4\textwidth}
         \centering
         \includegraphics[width=\textwidth]{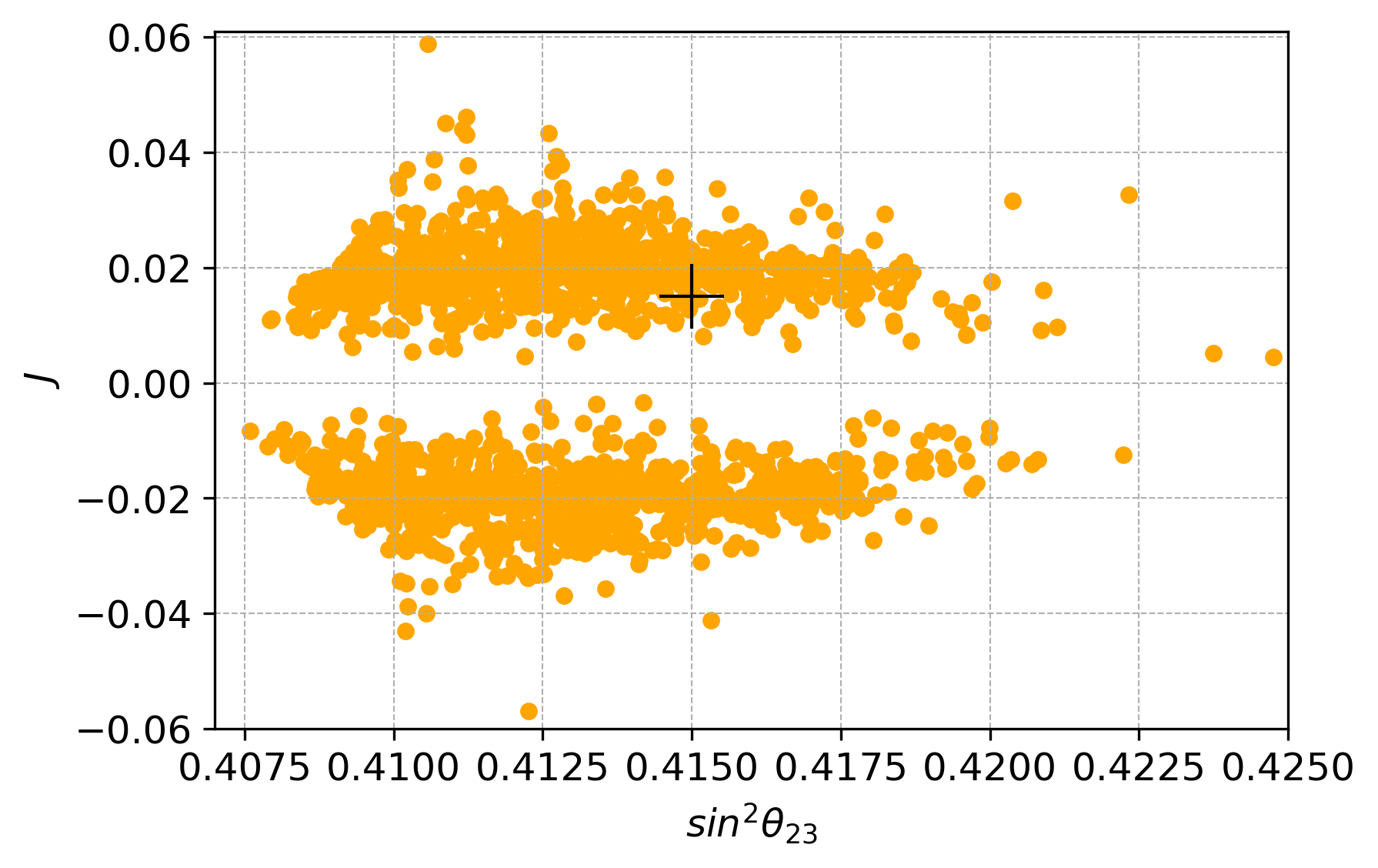}
     \end{subfigure}
     \hfill
     \begin{subfigure}[b]{0.4\textwidth}
         \centering
         \includegraphics[width=\textwidth]{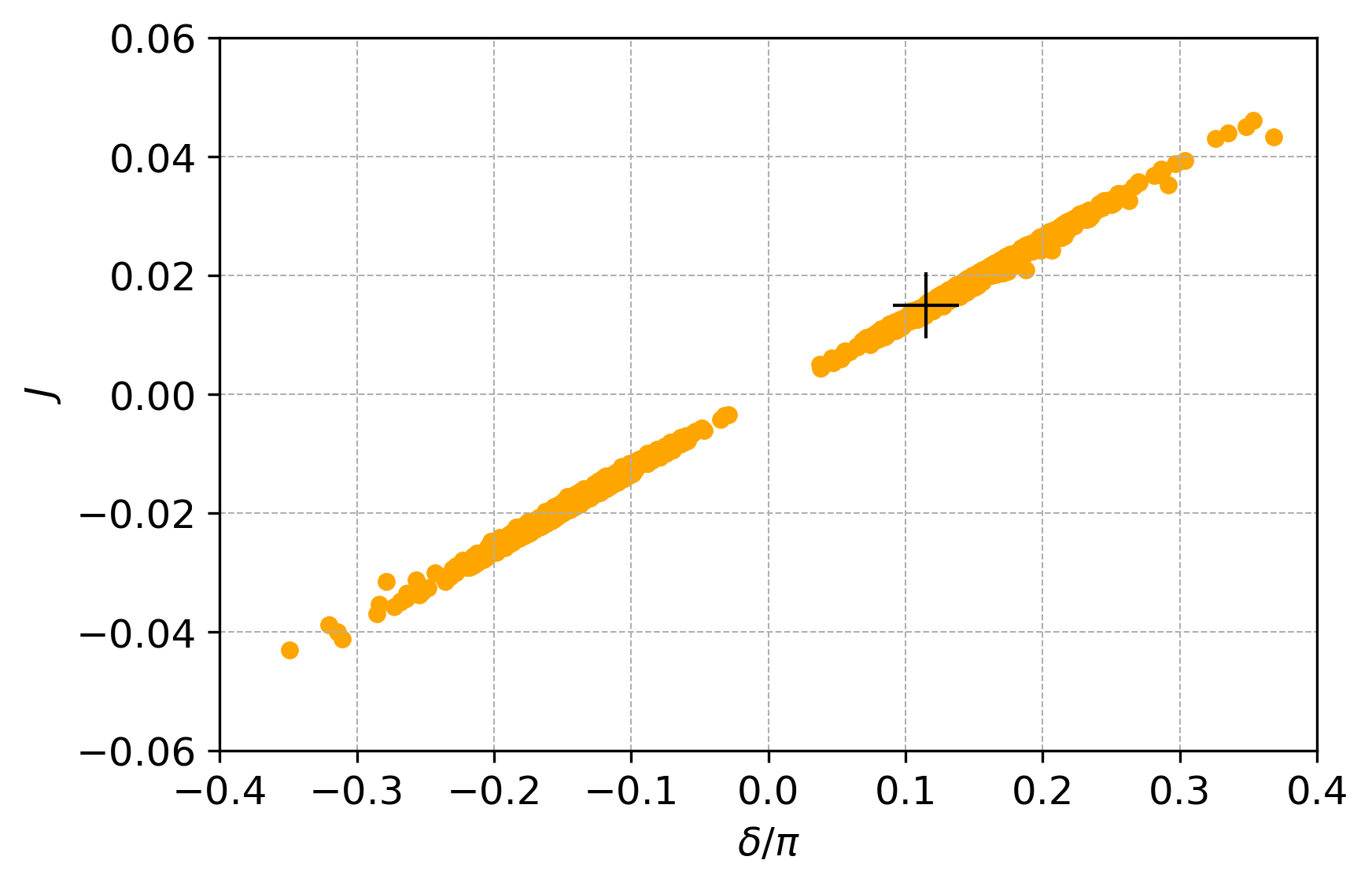}
     \end{subfigure}
    \caption{Correlation between the Jarlskog invariant parameter($J$) with Dirac CP($\delta_{CP}$) phase and atmospheric mixing angle($sin^2\theta_{23}$) respectively. The best fit value for 3$\sigma$ range is given by black +.}
    \label{fig:5}
\end{figure}

\noindent \textbf {Prediction for Jarlskog invariant Parameter:}
In Fig.\ref{fig:5} We are sharing further forecasts generated by the revised model concerning CP-violation's Jarlskog parameter($J$) and the effective Majorana mass($m_{ee}$) that defines the neutrinoless double beta decay. The Jarlskog constant is a quantity that remains unchanged even after a phase redefinition.
\begin{equation}
J= Im \{{U_{11}U_{22} U_{12}^{*} U_{21}^{*}}\} = s_{12}c_{13}^{2}s_{12}c_{12}s_{23}c_{23} sin\delta 
\end{equation}

\noindent \textbf {Neutrinoless double beta decay (NDBD):}
The NDBD phenomenon is crucial in the field of neutrino physics, as it involves the light Majorana neutrinos. The process is governed by an effective mass, denoted as $m_{ee}$, which can be calculated using the equation:
\begin{equation}
m_{ee}= U^2_{Li} m_i
\end{equation}
 where $U_{Li}$ are the elements of the first row of the neutrino mixing matrix $U_{PMNS}$. This equation depends on certain known parameters such as $\theta_{12}$ and $\theta_{13}$, as well as unknown Majorana phases denoted by $\alpha$ and $\beta$. The diagonalizing matrix of the light neutrino mass matrix, denoted by $m_\nu$, is represented by $U_{PMNS}$, such that 
 \begin{equation}
  m_\nu= U_{PMNS} M^{(diag)}_\nu U^T_{PMNS}
 \end{equation}
  where, $M^{(diag)}_\nu$ =diag($m_1$, $m_2$, $m_3$). The effective Majorana mass can be expressed applying the diagonalizing matrix elements and the mass eigenvalues as follows:
  \begin{equation}
     m_{ee}= m_1 c_{12}^2 c^2_{13}+ m_2 s^2_{12} c^2_{13} e^{2\iota \alpha} + m_3 s^2_{13} e^{2\iota \beta}
  \end{equation} where $c_{12}$ and $s_{12}$ are the cosine and sine of the mixing angle $\theta_{12}$, respectively.

\begin{figure}[!ht]
     \centering
     \begin{subfigure}[b]{0.46\textwidth}
         \centering
         \includegraphics[width=\textwidth,  height = 6cm]{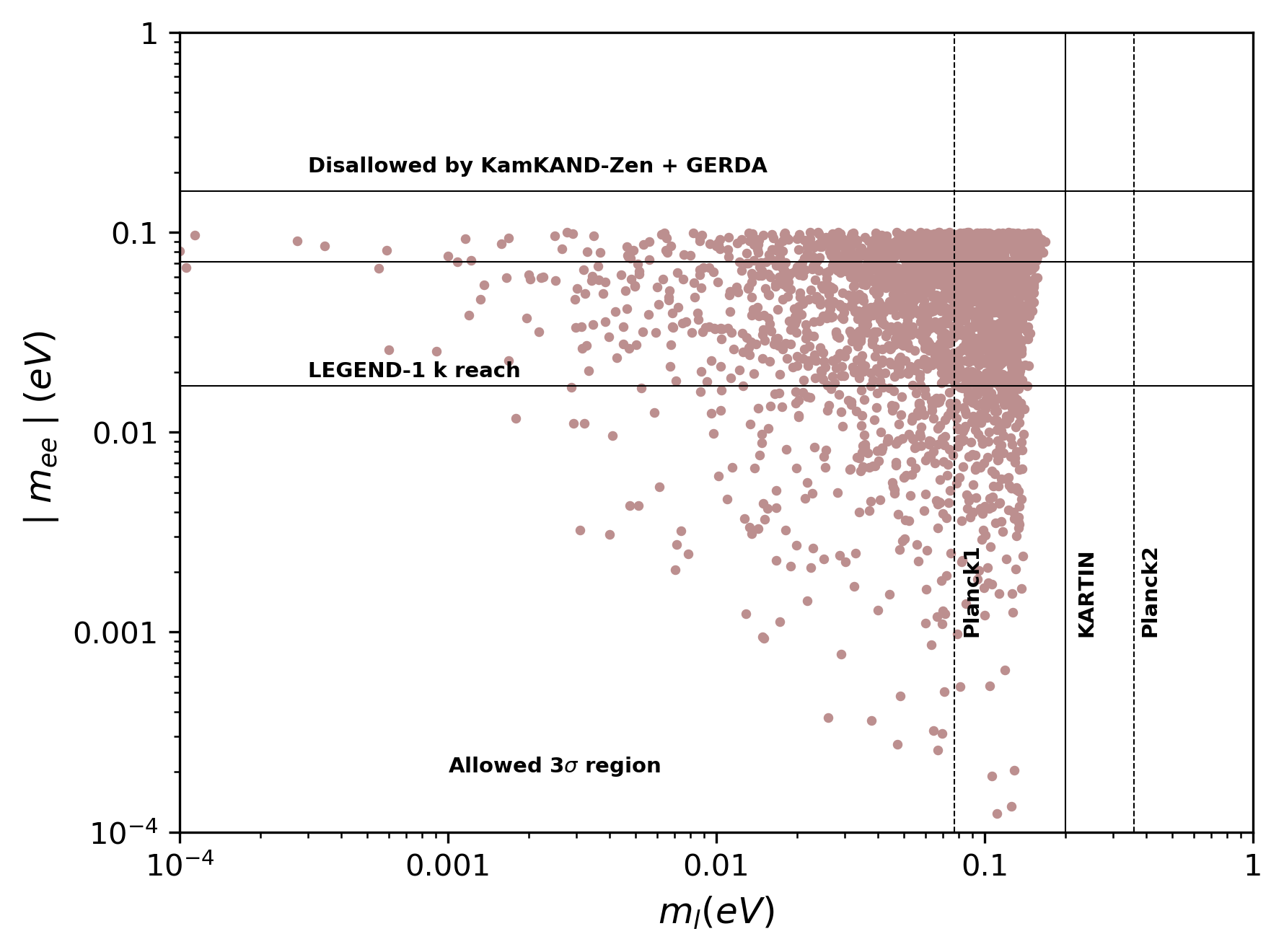}
     \end{subfigure}
 \hfill
     \begin{subfigure}[b]{0.46\textwidth}
         \centering
         \includegraphics[width=\textwidth,  height = 6cm]{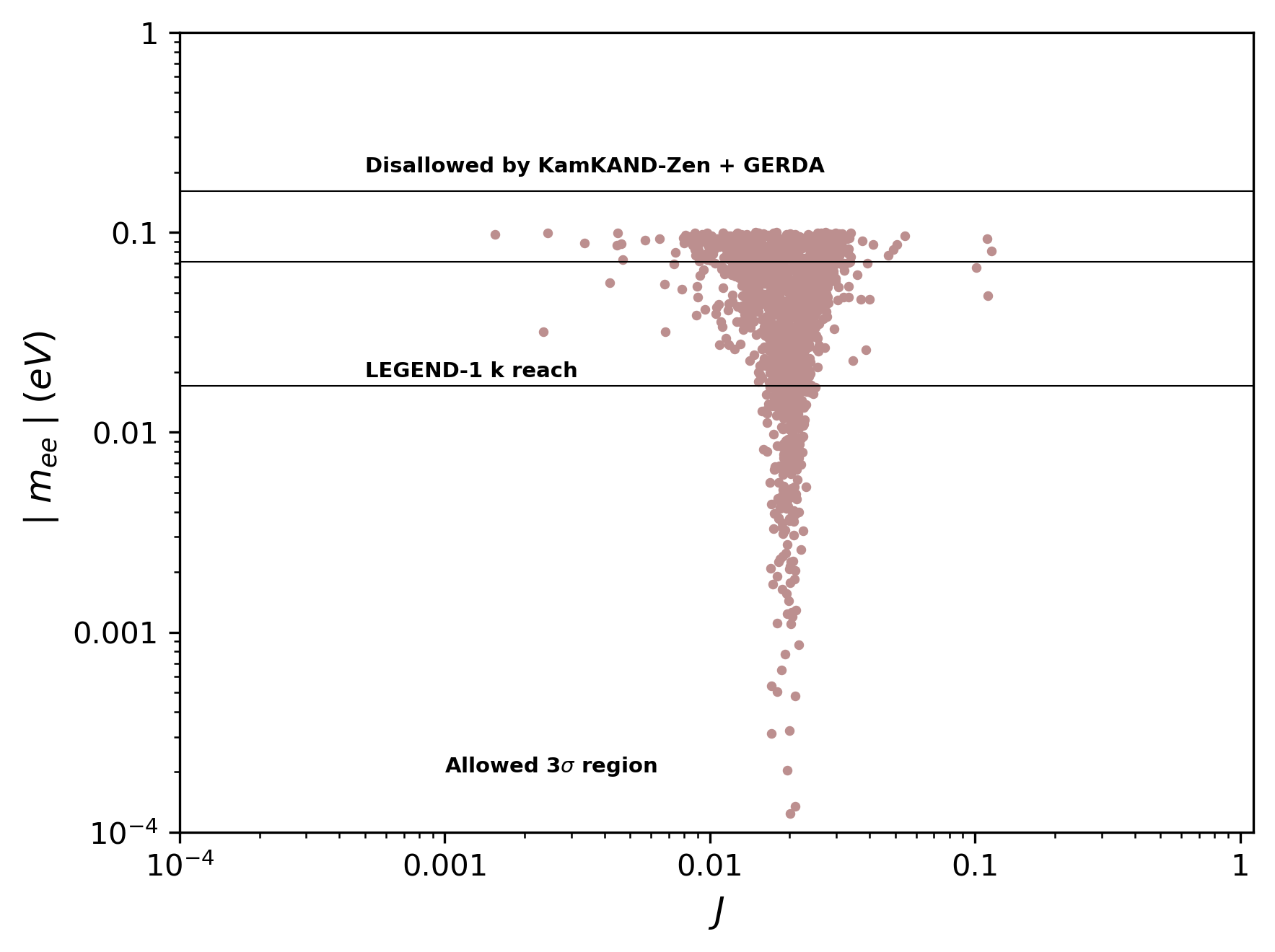}
     \end{subfigure}
  \caption{ Correlation between Effective Majorana neutrino
mass($m_{ee}$) with lightest neutrino mass($m_{1}$) and Jarlskog invariant($J$).}
 \label{fig:6}
\end{figure}

After analyzing the restricted parameter space, we have computed the value of $m_{ee}$ in the NH scenario. The figure labeled as \ref{fig:6} depicts the changes in $m_{ee}$ as per the lightest neutrino mass. Additionally, the sensitivity range of experiments like  GERDA, KamLAND-Zen and LEGEND-1k for neutrinoless double beta decay is also illustrated in the same figure. The combined constraints from KamLAND-Zen and GERDA experiments put an upper limit on $m_{ee}$ in the range 0.071–0.161 eV.  The vertical black solid line represents the future sentivity of KATRIN $m_{lightest} < 0.2 eV $\cite{katrin2001katrin} and the other two vertical lines represent the bound $m_{lightest} < 0.077 $eV
and $m_{lightest} < 0.36 $eV, following the two extreme bounds from Planck data set. As per the findings, $m_{ee}$ falls well within the reach of these NDBD experiments for normal hierarchy.

\section{Conclusion}
 We have introduced $\Delta(54)$ flavor model with two SM Higgs along with $Z_2 \otimes Z_3 \otimes Z_4$ symmetry  that generate a neutrino mass matrix.  We have constructed a flavor-symmetric approach in Inverse Seesaw mechanism to realize the neutrino masses and mixing to fit present neutrino oscillation data, including the non-zero reactor angle ($\theta_{13}$) and Dirac CP ($\delta_{CP}$). We use additional flavons to obtain the desired mixing pattern.  The model under consideration clearly shows deviation from TBM mixing of the neutrino mixing matrix. The predictions for the neutrino oscillation parameters derived from the resulting mass matrix are in agreement with the best-fit values obtained through $\chi^2$ analysis.  The prediction of the mixing angles, mass-squared differences, and the Dirac CP phase in the IH case do not agree with the experimental data. The prediction of mixing angles in NH case indicates that for the parameter space under consideration, the model prefers lower octant of $\theta_{23}$. \par
Furthermore, we investigated the Jarlskog invariant parameter and NDBD within the context of our $\Delta(54)$ models. The effective Majorana neutrino mass $|m_{ee}|$ is well within the sensitivity reach of the recent $0\nu \beta \beta$ experiments. In extension of this work, the model can be used to study leptogenesis and dark matter. The consistency of the model predictions with the latest neutrino data implies that model may be tested in future neutrino oscillation experiments. 

\section*{Acknowledgements} \par

HB acknowledges Tezpur University, India for Institutional Research Fellowship. The research of NKF is funded by DST-SERB, India under Grant no. EMR/2015/001683. AB acknowledges the CSIR, India for Senior Research Fellowship(file no 09/796(0072)/2017-
EMR-1). BT acknowledges the DST, India for INSPIRE Fellowship vide Grant no.
DST/INSPIRE/2018/IF180588. HB is thankful to Shawan Kumar Jha for fruitful discussions.

\section*{Appendix}

Tensor products of $\Delta(54)$

\begin{align*}
    1_{1}\otimes S_{i} = S_{i}, \qquad  1_{2}\otimes 1_{2} = 1_{1}, \qquad 1_{2}\otimes 3_{1(1)} = 3_{2(1)} \\
    1_{2}\otimes 3_{1(2)} = 3_{2(2)} , \qquad  1_{2}\otimes 3_{2(1)} = 3_{1(1)} , \qquad 1_{2}\otimes 3_{2(2)} = 3_{1(2)} 
\end{align*}

 \begin{align*}
    \begin{pmatrix}
    a_1\\ a_2\\a_3\end{pmatrix}_{3_{1(1)}} \otimes   \begin{pmatrix}
    b_1\\b_2\\b_3\end{pmatrix}_{3_{1(1)}} = \begin{pmatrix}
    a_{1}b_{1}\\a_{2}b_{2}\\a_{3}b_{3}\end{pmatrix}_{3_{1(2)}}  \oplus \begin{pmatrix}
    a_{2}b_{3} + a_{3}b_{2} \\a_{3}b_{1} + a_{1}b_{3} \\a_{1}b_{2} + a_{2}b_{1} \end{pmatrix}_{3_{1(2)}} \oplus  \begin{pmatrix}
    a_{2}b_{3} - a_{3}b_{2} \\a_{3}b_{1} - a_{1}b_{3} \\a_{1}b_{2} - a_{2}b_{1}\end{pmatrix}_{3_{2(2)}}   
\end{align*}

  \begin{align*}
    \begin{pmatrix}
    a_1\\ a_2\\a_3\end{pmatrix}_{3_{1(2)}} \otimes   \begin{pmatrix}
    b_1\\b_2\\b_3\end{pmatrix}_{3_{1(2)}} = \begin{pmatrix}
    a_{1}b_{1}\\a_{2}b_{2}\\a_{3}b_{3}\end{pmatrix}_{3_{1(1)}}  \oplus \begin{pmatrix}
    a_{2}b_{3} + a_{3}b_{2} \\a_{3}b_{1} + a_{1}b_{3} \\a_{1}b_{2} + a_{2}b_{1} \end{pmatrix}_{3_{1(1)}} \oplus  \begin{pmatrix}
    a_{2}b_{3} - a_{3}b_{2} \\a_{3}b_{1} - a_{1}b_{3} \\a_{1}b_{2} - a_{2}b_{1}\end{pmatrix}_{3_{2(1)}}   
\end{align*}

 \begin{align*}
    \begin{pmatrix}
    a_1\\ a_2\\a_3\end{pmatrix}_{3_{2(1)}} \otimes   \begin{pmatrix}
    b_1\\b_2\\b_3\end{pmatrix}_{3_{2(1)}} = \begin{pmatrix}
    a_{1}b_{1}\\a_{2}b_{2}\\a_{3}b_{3}\end{pmatrix}_{3_{1(2)}}  \oplus \begin{pmatrix}
    a_{2}b_{3} + a_{3}b_{2} \\a_{3}b_{1} + a_{1}b_{3} \\a_{1}b_{2} + a_{2}b_{1} \end{pmatrix}_{3_{1(2)}} \oplus  \begin{pmatrix}
    a_{2}b_{3} - a_{3}b_{2} \\a_{3}b_{1} - a_{1}b_{3} \\a_{1}b_{2} - a_{2}b_{1}\end{pmatrix}_{3_{2(2)}}   
\end{align*}

 \begin{align*}
    \begin{pmatrix}
    a_1\\ a_2\\a_3\end{pmatrix}_{3_{2(2)}} \otimes   \begin{pmatrix}
    b_1\\b_2\\b_3\end{pmatrix}_{3_{2(2)}} = \begin{pmatrix}
    a_{1}b_{1}\\a_{2}b_{2}\\a_{3}b_{3}\end{pmatrix}_{3_{1(1)}}  \oplus \begin{pmatrix}
    a_{2}b_{3} + a_{3}b_{2} \\a_{3}b_{1} + a_{1}b_{3} \\a_{1}b_{2} + a_{2}b_{1} \end{pmatrix}_{3_{1(1)}} \oplus  \begin{pmatrix}
    a_{2}b_{3} - a_{3}b_{2} \\a_{3}b_{1} - a_{1}b_{3} \\a_{1}b_{2} - a_{2}b_{1}\end{pmatrix}_{3_{2(1)}} 
\end{align*}

 \begin{align*}
     \begin{pmatrix}
    a_1\\ a_2\\a_3\end{pmatrix}_{3_{1(1)}} \otimes   \begin{pmatrix}
    b_1\\b_2\\b_3\end{pmatrix}_{3_{1(2)}} = &\begin{pmatrix}
    a_{1}b_{1} + a_{2}b_{2} + a_{3}b_{3} \end{pmatrix}_{1_{1}}  \oplus \begin{pmatrix}
    a_{1}b_{1} + \omega^2 a_{2}b_{2} + \omega a_{3}b_{3 }\\ \omega a_{1}b_{1} + \omega^2 a_{2}b_{2} +  a_{3}b_{3 }\end{pmatrix}_{2_{1}} \\&
    \oplus  \begin{pmatrix}
    a_{1}b_{2} + \omega^2 a_{2}b_{3} + \omega a_{3}b_{1 }\\\omega a_{1}b_{3} + \omega^2 a_{2}b_{1} + a_{3}b_{2}\end{pmatrix}_{2_{2}}  \oplus  \begin{pmatrix}
    a_{1}b_{3} + \omega^2 a_{2}b_{1} + \omega a_{3}b_{2}\\ \omega a_{1}b_{2} + \omega^2 a_{2}b_{3} + \omega a_{3}b_{1}\end{pmatrix}_{2_{3}} \\& \oplus \begin{pmatrix}
    a_{1}b_{3} +  a_{2}b_{1} + a_{3}b_{2}\\a_{1}b_{2} +  a_{2}b_{3} + a_{3}b_{1 }\end{pmatrix}_{2_{4}} \\
\end{align*}

\begin{align*}
    \begin{pmatrix}
    a_1\\ a_2\\a_3\end{pmatrix}_{3_{1(1)}} \otimes   \begin{pmatrix}
    b_1\\b_2\\b_3\end{pmatrix}_{3_{2(1)}} = \begin{pmatrix}
    a_{1}b_{1}\\a_{2}b_{2}\\a_{3}b_{3}\end{pmatrix}_{3_{2(2)}}  \oplus \begin{pmatrix}
    a_{3}b_{2} - a_{2}b_{3} \\a_{1}b_{3} - a_{3}b_{1} \\a_{2}b_{1} - a_{1}b_{2} \end{pmatrix}_{3_{1(2)}} \oplus  \begin{pmatrix}
    a_{3}b_{2} + a_{2}b_{3} \\a_{1}b_{3} + a_{3}b_{1} \\a_{2}b_{1} + a_{1}b_{2}\end{pmatrix}_{3_{2(2)}}   
\end{align*}

 \begin{align*}
     \begin{pmatrix}
    a_1\\ a_2\\a_3\end{pmatrix}_{3_{1(1)}} \otimes   \begin{pmatrix}
    b_1\\b_2\\b_3\end{pmatrix}_{3_{2(2)}} = &\begin{pmatrix}
    a_{1}b_{1} + a_{2}b_{2} + a_{3}b_{3} \end{pmatrix}_{1_{2}}  \oplus \begin{pmatrix}
    a_{1}b_{1} + \omega^2 a_{2}b_{2} + \omega a_{3}b_{3 }\\ -\omega a_{1}b_{1} - \omega^2 a_{2}b_{2} -  a_{3}b_{3 }\end{pmatrix}_{2_{1}} \\&
    \oplus  \begin{pmatrix}
    a_{1}b_{2} + \omega^2 a_{2}b_{3} + \omega a_{3}b_{1 }\\-\omega a_{1}b_{3} - \omega^2 a_{2}b_{1} - a_{3}b_{2}\end{pmatrix}_{2_{2}}  \oplus  \begin{pmatrix}
    a_{1}b_{3} + \omega^2 a_{2}b_{1} + \omega a_{3}b_{2}\\ -\omega a_{1}b_{2} - \omega^2 a_{2}b_{3} - a_{3}b_{1}\end{pmatrix}_{2_{3}} \\& \oplus \begin{pmatrix}
    a_{1}b_{3} +  a_{2}b_{1} + a_{3}b_{2}\\ -a_{1}b_{2} - a_{2}b_{3} -  a_{3}b_{1 }\end{pmatrix}_{2_{4}} \\
\end{align*}

 \begin{align*}
     \begin{pmatrix}
    a_1\\ a_2\\a_3\end{pmatrix}_{3_{1(2)}} \otimes   \begin{pmatrix}
    b_1\\b_2\\b_3\end{pmatrix}_{3_{2(1)}} = &\begin{pmatrix}
    a_{1}b_{1} + a_{2}b_{2} + a_{3}b_{3} \end{pmatrix}_{1_{2}}  \oplus \begin{pmatrix}
    a_{1}b_{1} + \omega^2 a_{2}b_{2} + \omega a_{3}b_{3 }\\ -\omega a_{1}b_{1} - \omega^2 a_{2}b_{2} -  a_{3}b_{3 }\end{pmatrix}_{2_{1}} \\&
    \oplus  \begin{pmatrix}
    a_{1}b_{2} + \omega^2 a_{2}b_{3} + \omega a_{3}b_{1 }\\-\omega a_{1}b_{3} - \omega^2 a_{2}b_{1} - a_{3}b_{2}\end{pmatrix}_{2_{2}}  \oplus  \begin{pmatrix}
    a_{1}b_{3} + \omega^2 a_{2}b_{1} + \omega a_{3}b_{2}\\ -\omega a_{1}b_{2} - \omega^2 a_{2}b_{3} - a_{3}b_{1}\end{pmatrix}_{2_{3}} \\& \oplus \begin{pmatrix}
    a_{1}b_{3} +  a_{2}b_{1} + a_{3}b_{2}\\ -a_{1}b_{2} - a_{2}b_{3} -  a_{3}b_{1 }\end{pmatrix}_{2_{4}} \\
\end{align*}

 \begin{align*}
    \begin{pmatrix}
    a_1\\ a_2\\a_3\end{pmatrix}_{3_{1(2)}} \otimes   \begin{pmatrix}
    b_1\\b_2\\b_3\end{pmatrix}_{3_{2(2)}} = \begin{pmatrix}
    a_{1}b_{1}\\a_{2}b_{2}\\a_{3}b_{3}\end{pmatrix}_{3_{2(1)}}  \oplus \begin{pmatrix}
    a_{3}b_{2} - a_{2}b_{3} \\a_{1}b_{3} - a_{3}b_{1} \\a_{2}b_{1} - a_{1}b_{2} \end{pmatrix}_{3_{1(1)}} \oplus  \begin{pmatrix}
    a_{3}b_{2} + a_{2}b_{3} \\a_{1}b_{3} + a_{3}b_{1} \\a_{2}b_{1} + a_{1}b_{2}\end{pmatrix}_{3_{2(1)}}   
\end{align*}

  \begin{align*}
     \begin{pmatrix}
    a_1\\ a_2\\a_3\end{pmatrix}_{3_{2(1)}} \otimes   \begin{pmatrix}
    b_1\\b_2\\b_3\end{pmatrix}_{3_{2(2)}} = &\begin{pmatrix}
    a_{1}b_{1} + a_{2}b_{2} + a_{3}b_{3} \end{pmatrix}_{1_{1}}  \oplus \begin{pmatrix}
    a_{1}b_{1} + \omega^2 a_{2}b_{2} + \omega a_{3}b_{3 }\\ \omega a_{1}b_{1} + \omega^2 a_{2}b_{2} +  a_{3}b_{3 }\end{pmatrix}_{2_{1}} \\&
    \oplus  \begin{pmatrix}
    a_{1}b_{2} + \omega^2 a_{2}b_{3} + \omega a_{3}b_{1 }\\ \omega a_{1}b_{3} + \omega^2 a_{2}b_{1} + a_{3}b_{2}\end{pmatrix}_{2_{2}}  \oplus  \begin{pmatrix}
    a_{1}b_{3} + \omega^2 a_{2}b_{1} + \omega a_{3}b_{2}\\ \omega a_{1}b_{2} + \omega^2 a_{2}b_{3} + a_{3}b_{1}\end{pmatrix}_{2_{3}} \\& \oplus \begin{pmatrix}
    a_{1}b_{3} +  a_{2}b_{1} + a_{3}b_{2}\\ a_{1}b_{2} + a_{2}b_{3} +  a_{3}b_{1 }\end{pmatrix}_{2_{4}} \\
\end{align*}

 \begin{align*}
    \begin{pmatrix}
    a_1\\ a_2 \end{pmatrix}_{2_{s}} \otimes   \begin{pmatrix}
    b_1\\ b_2 \end{pmatrix}_{2_{s}} = \begin{pmatrix}
    a_1 b_2 +  a_2 b_1 \end{pmatrix}_{1_{1}} \oplus \begin{pmatrix}
    a_1 b_2 -  a_2 b_1\end{pmatrix}_{1_{2}} \oplus \begin{pmatrix}
    a_2 b_2 \\ a_1 b_1\end{pmatrix}_{2_{s}}
\end{align*}

\begin{align*}
    \begin{pmatrix}
    a_1\\ a_2\end{pmatrix}_{2_{1}} \otimes   \begin{pmatrix}
    b_1\\b_2\end{pmatrix}_{2_{2}} = \begin{pmatrix}
    a_{2}b_{2}\\a_{1}b_{1}\end{pmatrix}_{2_{3}}  \oplus \begin{pmatrix}
     a_{2}b_{1}\\a_{1}b_{2} \end{pmatrix}_{2_{4}} 
\end{align*}

\begin{align*}
    \begin{pmatrix}
    a_1\\ a_2\end{pmatrix}_{2_{1}} \otimes   \begin{pmatrix}
    b_1\\b_2\end{pmatrix}_{2_{3}} = \begin{pmatrix}
    a_{2}b_{2}\\a_{1}b_{1}\end{pmatrix}_{2_{2}}  \oplus \begin{pmatrix}
     a_{2}b_{1}\\a_{1}b_{2} \end{pmatrix}_{2_{4}} 
\end{align*}

\begin{align*}
    \begin{pmatrix}
    a_1\\ a_2\end{pmatrix}_{2_{1}} \otimes   \begin{pmatrix}
    b_1\\b_2\end{pmatrix}_{2_{4}} = \begin{pmatrix}
    a_{1}b_{2}\\a_{2}b_{1}\end{pmatrix}_{2_{2}}  \oplus \begin{pmatrix}
     a_{1}b_{1}\\a_{2}b_{2} \end{pmatrix}_{2_{3}} 
\end{align*}

\begin{align*}
    \begin{pmatrix}
    a_1\\ a_2\end{pmatrix}_{2_{2}} \otimes   \begin{pmatrix}
    b_1\\b_2\end{pmatrix}_{2_{3}} = \begin{pmatrix}
    a_{2}b_{2}\\a_{1}b_{1}\end{pmatrix}_{2_{1}}  \oplus \begin{pmatrix}
     a_{1}b_{2}\\a_{2}b_{1} \end{pmatrix}_{2_{4}} 
\end{align*}

\begin{align*}
    \begin{pmatrix}
    a_1\\ a_2\end{pmatrix}_{2_{2}} \otimes   \begin{pmatrix}
    b_1\\b_2\end{pmatrix}_{2_{4}} = \begin{pmatrix}
    a_{1}b_{1}\\a_{2}b_{2}\end{pmatrix}_{2_{1}}  \oplus \begin{pmatrix}
     a_{1}b_{2}\\a_{2}b_{1} \end{pmatrix}_{2_{3}} 
\end{align*}

\begin{align*}
    \begin{pmatrix}
    a_1\\ a_2\end{pmatrix}_{2_{3}} \otimes   \begin{pmatrix}
    b_1\\b_2\end{pmatrix}_{2_{4}} = \begin{pmatrix}
    a_{1}b_{2}\\a_{2}b_{1}\end{pmatrix}_{2_{1}}  \oplus \begin{pmatrix}
     a_{1}b_{1}\\a_{2}b_{2} \end{pmatrix}_{2_{2}} 
\end{align*}

\bibliographystyle{unsrt}
\bibliography{references}

\end{document}